\documentclass[journal]{IEEEtran}
\usepackage{amsmath,amsfonts}
\usepackage{algorithmic}
\usepackage{algorithm}
\usepackage{array}
\usepackage[caption=false,font=normalsize,labelfont=sf,textfont=sf]{subfig}
\usepackage{textcomp}
\usepackage{stfloats}
\usepackage{url}
\usepackage{verbatim}
\usepackage{graphicx}
\usepackage{cite}
\usepackage{pgfplots}  
\pgfplotsset{compat=1.17} 
\usepackage{etoolbox}
\usepackage{mathrsfs} 
\usepackage{multirow}
\usepackage{amsmath}
\usepackage{amsthm}

\begin{document}

\title{Mobility-Aware Multi-Task Decentralized Federated Learning for Vehicular Networks: Modeling, Analysis, and Optimization}

\author{
Dongyu Chen, Tao Deng, He Huang, Juncheng Jia, Mianxiong~Dong, Di Yuan, and Keqin Li,~\IEEEmembership{Fellow,~IEEE}

\thanks{D. Chen, T.\ Deng, H.\ Huang, and J.\ Jia are with School of Computer Science and Technology, Soochow University, Suzhou, Jiangsu 215006, China (E-mail:  dychan2000@gmail.com, \{dengtao;~huangh;~jiajuncheng\}@suda.edu.cn.)

M. Dong is with Department of Sciences and Informatics, Muroran Institute of Technology, Muroran 050-8585, Japan (E-mail: mx.dong@csse.muroran-it.ac.jp)

D.\ Yuan is with the Department of Information Technology, Uppsala University, 751 05 Uppsala, Sweden (E-mail: di.yuan@it.uu.se)

K.\ Li is with the Department of Computer Science, State University of New York, New Paltz, New York 12561, USA (E-mail:
lik@newpaltz.edu)
}
}



\maketitle

\begin{abstract}
Federated learning (FL) is a promising paradigm that 
can enable collaborative model training between vehicles while protecting data privacy, thereby significantly improving the performance of intelligent transportation systems (ITSs).
In vehicular networks, due to mobility, resource constraints, and the concurrent execution of multiple training tasks, how to allocate limited resources effectively to achieve optimal model training of multiple tasks is an extremely challenging issue.
In this paper, we propose a mobility-aware multi-task decentralized federated learning (MMFL) framework for vehicular networks.
By this framework, we address task scheduling, subcarrier allocation, and leader selection, as a joint optimization problem, termed as TSLP.
For the case with a single FL task, we derive the convergence bound of model training. 
For general cases, we first model TSLP as a resource allocation game, and prove the existence of a Nash equilibrium (NE).
Then, based on this proof, we reformulate the game as a decentralized partially observable Markov decision process (DEC-POMDP), and develop an algorithm based on heterogeneous-agent proximal policy optimization (HAPPO) to solve DEC-POMDP.
Finally, numerical results are used to demonstrate the effectiveness of the proposed algorithm. 
\end{abstract}


\section{Introduction}

\IEEEPARstart{W}{ith} the rapid advancement of intelligent vehicles (IVs), the intelligent transportation system (ITS) significantly enhances travel experience and safety \cite{yan2024edge}.
Deep learning technologies have enabled various neural network models to offer driving assistance features such as traffic flow prediction (TFP), free-space detection (FSD), and driving behavior monitoring (DBM), thus enhancing the overall driving experience \cite{zhang2023federated}.
Traditional deep learning methods, which rely on centralized data aggregation and training, face challenges related to communication pressure and data privacy issues. 
To address these issues, 
federated learning (FL) has been introduced for ITS \cite{McMahan2017}.
FL involves multiple vehicles and infrastructures, where each participating vehicle has its own local dataset for training.
The aggregator, vehicle or infrastructure,  collects local model updates from the vehicles.
This approach shows great potential in improving both the efficiency and privacy of deep learning in vehicular networks.

In vehicular networks, there are numerous model training demands due to the following two aspects.
First, IVs must manage multiple tasks concurrently. For example, during a journey, vehicles may employ 3D object detection for road condition monitoring, DBM for trajectory planning based on the positions of other vehicles, and TFP for real-time navigation utilizing traffic incident data \cite{deng2023review}.
Second, even within a single task, there may exist diversity in the need of model training. 
For instance, the work in \cite{zeng2021multi} 
 utilizes spatiotemporal dependencies for clustering and employs FL to train multiple path planning models.
Therefore, addressing the multi-task FL challenge in vehicular networks is of urgent technical importance.
However, implementing multi-task FL presents a substantial challenge, as it must address the impact of vehicle mobility, reduce latency, and optimize energy efficiency.
Existing models are limited in two respects.
On one hand, they primarily focus on FL for single-task scenarios in vehicular networks, making it hard to be used for multi-task scenarios.
They need to address issues such as single point of failure and overwriting model update \cite{Savazzi2020DFLIOT}.
On the other hand, while some models are trained to handle multiple tasks, careful consideration of the affinity between tasks is necessary \cite{zhuang2023mas}.
A lack of task affinity can lead to negative transfer, which degrades training performance.
Furthermore, these models often overlook the impact of client mobility and the resource constraints, and both exist in vehicular networks.



We explore vehicle mobility in vehicular networks to develop optimization strategies for multi-task FL while efficiently utilizing scarce resources.
Our goal is to minimize the model training loss of each task.
The main contributions of this paper are as follow:
\begin{enumerate}
    \item We design a mobility-aware multi-task decentralized federated learning (MMFL) framework for vehicular networks.
    The MMFL framework adopts a multi-leader-multi-follower approach for multi-task scenarios.
    Vehicles that are close to each other communicate directly using vehicle-to-vehicle (V2V) communication, while vehicles that are farther apart communicate indirectly using vehicle-to-infrastructure (V2I) communication.
    \item To improve the training efficiency of MMFL, we address task scheduling, subcarrier allocation, and leader selection, as a joint optimization problem, termed as TSLP, taking into account vehicle mobility, resource constraints, as well as latency constrains.
    \item For problem solving, we analyze the convergence bound of model training for 
    the single-task scenario under the MMFL framework. 
    For general scenarios, we first model TSLP as a resource allocation game among multiple tasks. We prove the existence of a Nash equilibrium (NE) for the game.
    Subsequently, we reformulate the game as a decentralized partially observable Markov decision process (DEC-POMDP).
    Finally, we design an algorithm based on Heterogeneous-Agent Proximal Policy Optimization (HAPPO) to solve DEC-POMDP.
    \item We design a series of experiments to validate the effectiveness of the proposed algorithm.
    We utilize the urban mobility (SUMO) simulation software to generate traffic flow and simulate multi-task FL on four datasets, MNIST, FashionMNIST, SVHN, and CIFAR-10.
    We compare the proposed algorithm with baseline algorithms under various conditions, including varying vehicle initial energy, number of vehicles, number of tasks, and indirect transmission costs.
\end{enumerate}

\section{Related Work}

\subsection{Federated Learning in Vehicular Networks}

In vehicular networks, FL can be categorized into two types: centralized FL (CFL) and decentralized FL (DFL).


CFL relies on a central server for aggregating client models, requiring extra physical nodes as aggregation centers.
Some works use RSUs as a model aggregation node \cite{Xie2022MOBFL, zhang2024vehicle,zhang2024joint,Zhang2023PALORA,yan2024dynamic,Pervej2023Resource,Singh2024DRL-Based,chen2024mobilityacc,Xiang2024Collaborative,Macedo2023Multiple}.
The works in \cite{Xie2022MOBFL, zhang2024vehicle, zhang2024joint} improve resource utilization during the vehicles' sojourn period by optimizing round duration, client selection, and compression ratio, thereby accelerating the convergence speed.
The work in \cite{Zhang2023PALORA} uses real-world data to model a two-dimensional scenario and avoids model unfairness by selecting appropriate devices to participate in the training process.
The work in \cite{yan2024dynamic} uses V2V communication to assist V2I communication, improving the transmission success rate during the sojourn period.
The work in \cite{Pervej2023Resource} assigns aggregation weights to vehicles based on their sojourn period at the RSU.
The work in \cite{Singh2024DRL-Based} assigns higher weights to vehicles with overdue model uploads and discards failed uploads.
The work in \cite{chen2024mobilityacc} discusses the convergence of mobility-aware FL under a three-layer network architecture. It proves that mobility is beneficial for the convergence of FL.
The work in \cite{Xiang2024Collaborative} dynamically adjusts the number of local iterations in a three-tier network architecture, effectively reducing the number of communication rounds.
The work in \cite{Macedo2023Multiple} ignores the update requests from vehicles that are about to leave the RSU coverage area.
However, high-speed vehicles have only short in-contact periods with RSUs, resulting in insufficient use of data.
Moreover, an overreliance on static RSUs poses the risk of a single point of failure \cite{Savazzi2020DFLIOT}.

DFL dynamically assigns model aggregation tasks to clients, eliminating the need for a fixed central server and enabling broader training through vehicular ad-hoc networks (VANETs) without reliance on fixed-location RSUs. This dynamic selection of aggregation nodes also mitigates the risk of single-point-of-failure.
DFL is divided into fully DFL and leader-follower DFL.
In fully DFL \cite{Hu2019DFL}, all clients act as leaders to perform aggregation tasks, and there is no clear hierarchical structure among the clients.
The work in \cite{liu2022enhance} assumes that all vehicles upload their local models to the nearest RSU, where fully DFL is performed between the RSUs.
In contrast, leader-follower DFL maintains a hierarchical model where the aggregation nodes centrally handle the aggregation tasks
\cite{AbdulRahman2023,Tan2024SC1BCS,Yang2022LeadFN,behera2021federated,Gou2024Vote,zhang2024leader}.
In \cite{AbdulRahman2023}, clusters are formed based on a QoS scoring system, with leaders conducting preliminary aggregation before final consolidation at the RSU.
The work in \cite{Tan2024SC1BCS} uses soft clustering to form clusters and applies compression to reduce the transmission burden for duplicate models.
The work in \cite{Yang2022LeadFN} selects leaders based on computational, communication, and energy supply capabilities, ultimately enabling effective management of model aggregation and accelerating the FL process.
The work in \cite{behera2021federated} designs a Raft consensus mechanism for electing leaders.
The work in \cite{Gou2024Vote} uses mathematical derivation to further quantify the communication efficiency, the accuracy of joint decision-making, and the probability of successful leader election.
For DFL on satellites, the work in \cite{zhang2024leader} jointly optimizes leader selection and satellite communication resources.
However, this approach presents a challenge, as vehicles with outdated models acting as leaders can overwrite recent updates, leading to inefficient energy use and reduced training effectiveness, especially when multiple training tasks are present concurrently.


\subsection{Multi-task Federated Learning}

In multi-task FL scenarios, some existing works train a model that can perform multiple tasks \cite{sun2020adashare,yu2020gradient,zhuang2023mas}.
The work in \cite{sun2020adashare} uses soft parameter sharing to achieve optimal recognition accuracy while considering resource efficiency.
The work in \cite{yu2020gradient} employs neural architecture search for projecting task gradients onto orthogonal planes relative to conflicting gradients, thereby mitigating gradient conflicts.
The work in \cite{zhuang2023mas} first performs joint training, and then splits the model for separate training to alleviate the negative transfer issue between tasks.
The works in \cite{fu2023efficient,shi2024fairness,zhou2022efficient,yang2024efficient,cao2024efficient,li2024joint,wei2024joint} assign multiple FL tasks to different devices for training, but they do not consider the impact of mobility in vehicular networks.
The work in \cite{fu2023efficient} designs a client-sharing mechanism across multiple tasks, achieving a performance improvement of up to 2.03×.
The work in \cite{zhou2022efficient} addresses the scenario where multiple FL servers simultaneously attempt to select clients from the same pool. Based on Bayesian optimization, it jointly considers training time and the importance of preventing long waiting times.
The work in \cite{shi2024fairness} solves the waiting issue based on Lyapunov optimization.
The work in \cite{yang2024efficient} implements multi-task learning in the vertical FL scenario based on the rolling horizon method.
The work in \cite{cao2024efficient} delivers superior performance of multi-task FL for edge computing by leveraging the block coordinate descent algorithm.
The work in \cite{li2024joint} jointly optimizes task scheduling and resource allocation by considering the triple heterogeneity of data, devices, and tasks in a wireless network framework.
The work in \cite{wei2024joint} shows that FL models with adaptive topologies have lower learning costs, and subsequently employs a heuristic algorithm to minimize the total cost.


\section{System Model}

\subsection{MMFL Framework}

\begin{figure*}[!t]
  \centering
  \includegraphics[width=0.9\textwidth]{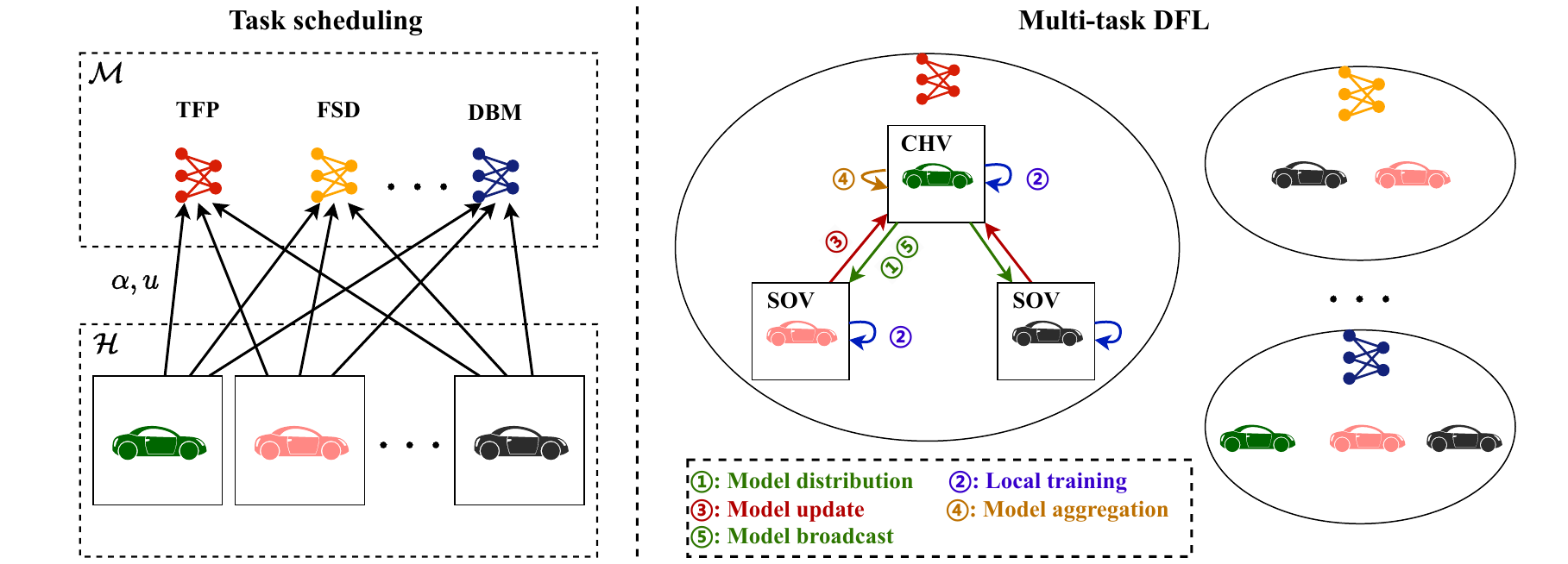}\\
  \caption{Workflow of one communication round with task scheduling followed by multi-task DFL.}
  \label{fig:workflow}
\end{figure*}
Fig. \ref{fig:workflow} shows the MMFL framework that includes $H$ vehicles and theses vehicles collaboratively complete $M$ training tasks, such as TFP, FSD, and DBM \cite{yan2024edge}.
Denote by $\mathcal{H}$ and $\mathcal{M}$ the sets of vehicles and tasks, respectively.
Denote by $\mathcal{W}_h=\{\boldsymbol{w}^1_h,\boldsymbol{w}^2_h,...,\boldsymbol{w}^M_h\}$ the set of model parameters for vehicle $h$, where $\boldsymbol{w}^m_h$ (for $m=1,\dots,M$) correspond one-by-one with the elements in the training task set $\mathcal{M}$.
Denote by $\mathcal{K}$ the set of communication rounds, $\mathcal{K}=\{1,2,...,K\}$.
All communication rounds are of the same maximum time duration $t^{round}$.
In the $k$-th communication round, the two-dimensional coordinates of vehicle $h$ are denoted by $(x_{kh},y_{kh})$, and its speed is denoted by  $s_{kh}$, ${\forall}h\in\mathcal{H}, {\forall}k\in\mathcal{K}$.
We assume that each vehicle can participate in at most one task training during a communication round. 
Specifically, denote by $\alpha^m_{kh}$ a binary variable, $\alpha^m_{kh} \in \{0,1\}$, with $\sum_{m\in\mathcal{M}}\alpha^m_{kh}\leq1, {\forall}h\in\mathcal{H}, {\forall}k\in\mathcal{K}$, which is one if and only if vehicle $h$ selects to train task $m$ in the $k$-th communication round. 
Denote by $\boldsymbol{\alpha}$ the corresponding tensor, where $\boldsymbol{\alpha}_k$ is a matrix containing the task scheduling status for each task of all vehicles in the $k$-th communication round.
The model training process employs a leader-follower DFL strategy \cite{zhang2024leader}. 
We divide the vehicles into two categories, i.e., cluster head vehicles (CHVs) and source vehicles (SOVs).
Each CHV corresponds to a task.
The set of SOVs is divided into $M$ subsets, each responding to a CHV.
Since we consider a dynamic cluster head strategy, the elements in these two sets will change dynamically in each communication round.
Denote by $u^m_{kh}$ a binary variable, $u^m_{kh}\in\{0,1\}$, with $\sum_{h\in\mathcal{H}}\alpha^m_{kh}u^m_{kh}=1,{\forall}m\in\mathcal{M},{\forall}k\in\mathcal{K}$, and $u^m_{kh} \leq \alpha^m_{kh},{\forall}m\in\mathcal{M},{\forall}k\in\mathcal{K},{\forall}h\in\mathcal{H}$, which is one if and only if vehicle $h$ is selected as the leader for task $m$ in the $k$-th communication round, and also serves as the CHV for task $m$.
Denote by $\boldsymbol{u}$ the corresponding tensor, where $\boldsymbol{u}_k$ is a matrix containing the leader selection status for each task of all vehicles in the $k$-th communication round, $\boldsymbol{u}^m_k$ is a vector containing the leader selection status for task $m$ of all vehicles in the $k$-th communication round.

\subsection{MMFL Training Model}

For task $m$, the training process consists of the following five parts.

\begin{enumerate}
    \item \textbf{Model distribution}: At the beginning of the $k$-th communication round, denote by $r$ the CHV. This CHV sends its local model, denoted by $\boldsymbol{w}^m_{k-1,r}$, to the SOVs.
    Furthermore, this model $\boldsymbol{w}^m_{k-1,r}$ also serves as the global model for task $m$, so we simplify $\boldsymbol{w}^m_{k-1,r}$ to $\boldsymbol{w}^m_{k-1}$. The local model of an SOV $s$, denoted by $\boldsymbol{w}^m_{k-1,s}$, is as follows by model distribution.
    \begin{equation}
        \boldsymbol{w}^m_{k-1,s}=\boldsymbol{w}^m_{k-1}=\boldsymbol{w}^m_{k-1,r}.
        \label{eq:Initial Model}
    \end{equation}
    \item \textbf{Local training}: After receiving the global model $\boldsymbol{w}^m_{k-1}$, each vehicle $h \in \mathcal{H}$ with $\alpha^m_{kh} = 1$ uses the stochastic gradient descent (SGD) algorithm to update the local model.
    Denote by $\mathcal{X}^m_h$ the data samples used for training by vehicle $h$ for task $m$.
    Denote by $\mathcal{Y}^m_h$ the data samples used for testing by vehicle $h$ for task $m$.
    Each training dataset has an associated distribution $\mathcal{D}^m_h$ over the space of samples $\mathcal{X}^m_h$, while each test dataset has an associated distribution $\mathcal{D}^{m'}_h$ over the space of samples $\mathcal{Y}^m_h$.
    The local training process is expressed as
    \begin{equation}
        \boldsymbol{w}^m_{kh}=\boldsymbol{w}^m_{k-1}-\frac{\eta_{k}}{B_{k}}\sum_{x\in\mathcal{B}_{kh}}\nabla f^m(\boldsymbol{w}^m_{k-1,h};\boldsymbol{x}),
        \label{eq:loss-define}
    \end{equation}
    where $\eta_k$ represents the learning rate of the $k$-th communication round, $\nabla f^m(\boldsymbol{w}^m_{k-1,h};\boldsymbol{x})$ represents the gradient of the local model loss function, and $\mathcal{B}_{kh}$ represents the subset of $\mathcal{X}^m_h$ generated by vehicle $h$ based on task $m$ during the $k$-th communication round. Assuming all vehicles have equal batch sizes, we have $|\mathcal{B}_{kh}| = B_k$.
    \item \textbf{Model update}: Each SOV $s$ sends its local model $\boldsymbol{w}^m_{ks}$ to the CHV.
    \item \textbf{Model aggregation}: The CHV aggregates all the received models by using a weighted average method,
    \begin{equation}
        w^m_{k}=
                \frac
                {\Sigma_{h\in\mathcal{H}}\alpha^m_{kh}|\mathcal{D}^m_h|\boldsymbol{w}^m_{kh}}
                {\sum_{h\in\mathcal{H}}\alpha^m_{kh}|\mathcal{D}^m_h|}.
        \label{eq:aggregation-define}
    \end{equation}
    \item \textbf{Model broadcast}: The CHV sends the aggregated model $\boldsymbol{w}^m_{k}$ to the SOVs.
    
\end{enumerate}
The above five training steps end until the loss function of each task converges or the number of communication rounds reaches its upper bound.

Each vehicle $h \in \mathcal{H}$ holds $M$ local datasets corresponding to the $M$ tasks.
Denote by $I$ the local iterations.
In each communication round, each vehicle performs $I$ local iterations.
In each training iteration, the vehicle extracts a data sample $\boldsymbol{x}$,  $\boldsymbol{x}\in\mathcal{X}^m_{h}$, from the dataset for training, and a data sample $\boldsymbol{y}$,  $\boldsymbol{y}\in\mathcal{Y}^m_{h}$, from the test dataset for evaluation.
The loss function of task $m$ in vehicle $h$, denoted by $f^m_{h}(\boldsymbol{w}^m)$, is expressed as
\begin{equation}
    f^m_{h}(\boldsymbol{w}^m) = \underset{\boldsymbol{x}\sim\mathcal{D}^m_{h}}{\operatorname*{\mathbb{E}}}[f^m(\boldsymbol{w}^m;\boldsymbol{x})],
    \label{eq:loss}
\end{equation}
where the loss function $f^m(\boldsymbol{w}^m; \boldsymbol{x})$ is used to measure the fitting performance of the model parameter $\boldsymbol{w}^m$.
In (\ref{eq:loss}), $f^m_{h}(\boldsymbol{w}^m)$ describes the average loss over the distribution $\mathcal{D}^m_h$.
We assume that the vehicles participating in task $m$ are drawn from the given distribution $\mathcal{P}$ \cite{yan2024dynamic}.
The global loss function of task $m$, denoted by $F^m(\boldsymbol{w}^m)$, is expressed as
\begin{equation}
    F^m(\boldsymbol{w}^m) = \underset{h\sim\mathcal{P}}{\operatorname*{\mathbb{E}}}[f^m_{h}(\boldsymbol{w}^m)].
    \label{eq:global-loss}
\end{equation}

\subsection{Mobility Model}

Due to mobility, the communication state of vehicles changes rapidly with their position.
For the convenience of analysis, we make a quasi-static assumption where a vehicle's position does not change during the communication and computation process.
Denote by $d_{kij}$ the distance between vehicle $i$ and vehicle $j$ for the $k$-th communication round, which is expressed as
\begin{equation}
    d_{kij}=\sqrt{(x_{ki}-x_{kj})^2+(y_{ki}-y_{kj})^2}.
    \label{eq:distance}
\end{equation}

\begin{figure}[!htbp]
  \centering
  \includegraphics[width=0.48\textwidth]{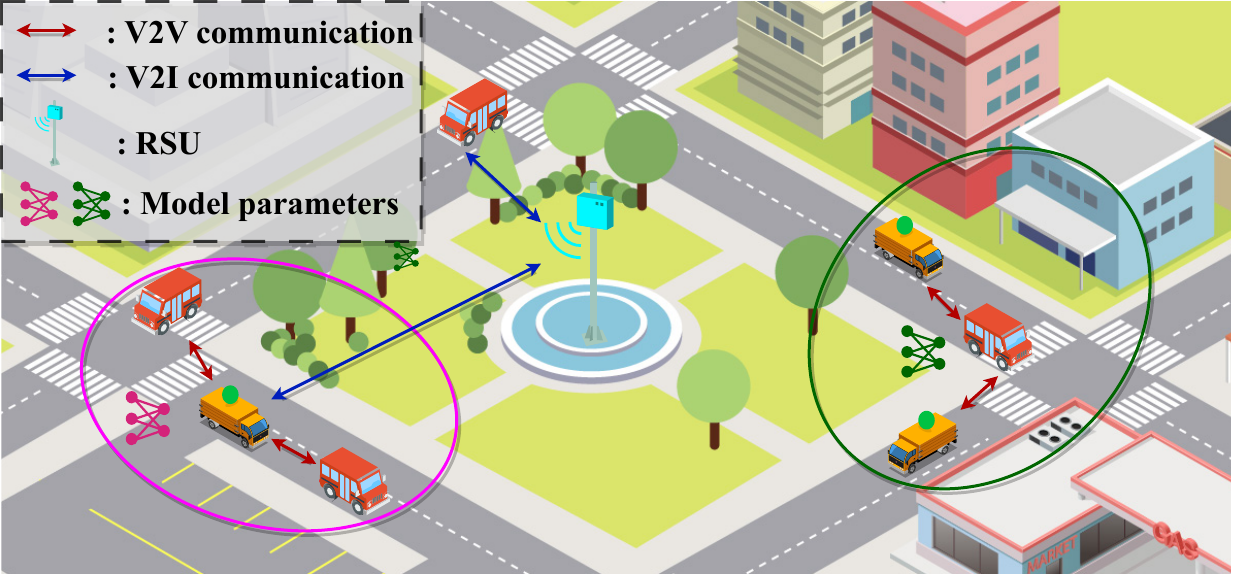}\\
  \caption{Network architecture.}
  \label{fig:network}
\end{figure}

As shown in Fig. \ref{fig:network}, we adopt a hybrid communication mode for model distribution, update, and broadcast.
We assume that the time and energy consumption for data reception and transmission are equal.
Specifically, denote by $d^U$ the direct communication radius.
When $d_{kij} \leq d^U$, vehicle $i$ and vehicle $j$ use V2V communication.
Conversely, when $d_{kij} > d^U$, the communication between vehicle $i$ and vehicle $j$ involves indirect communication.

\subsection{Communication Model}

We consider an orthogonal frequency division multiple access (OFDMA) as the transmission scheme \cite{zhang2024joint}.
This technology divides the bandwidth into multiple subcarriers, which are then allocated to different users.
The subcarrier resources are also orthogonally allocated between each CHV.
Denote by $l_{kij}$ the bandwidth allocation ratio assigned to SOV $s$ by CHV $r$ in the $k$-th communication round, which is expressed as
\begin{equation}
    l_{ksr}=\frac{1}{N} \sum_{n=1}^{N} \sum_{m \in \mathcal{M}} u^m_{kr} c^m_{ksn},
    \label{eq:bandwidth-allocation}
\end{equation}
where $N$ represents the number of subcarriers with the condition that $N \geq H$, and $c^m_{ksn}$ represents a binary variable, $c^m_{ksn}\in\{0,1\}$, with $\sum_{s \in \mathcal{H}}c^m_{ksn} \leq 1, \forall n \leq N, \forall m \in \mathcal{M},  \forall k \in \mathcal{K}$, which is one if and only if the $n$-th subcarrier is allocated to SOV $s$ in the $k$-th communication round for the $m$-th task.
Denote by $\boldsymbol{c}$ the corresponding tensor, where $\boldsymbol{c}_k$ is a tensor containing the bandwidth allocation status in the $k$-th communication round, and $\boldsymbol{c}^m_k$ is a matrix containing the bandwidth allocation status of task $m$ in the $k$-th communication round.
Denote by $R_{ksr}$ the transmission rate from SOV $s$ to CHV $r$ during the $k$-th communication round.
Based on the definition in \cite{zhang2024joint}, $R_{ksr}$ is expressed as
\begin{equation}
    R_{ksr}
    =l_{ksr} W \log _2
    \left (
    1 + \frac{p_{ks} h_{ksr} d_{ksr}^{-\nu}}{\sigma^2}
    \right ),
    \label{eq:transmission-rate}
\end{equation}
where $W$ represents the total uplink bandwidth, $p_{ks}$ represents the uplink transmission power of SOV $s$ in the $k$-th communication round, $h_{ksr}$ represents the power gain of the channel at a reference distance of 1 m in the $k$-th communication round, $\nu$ represents the path loss exponent, and $\sigma^2$ represents the variance of the complex white Gaussian channel noise.
For indirect communication, $d$ takes a large fixed value $\xi \cdot d^U$, where $\xi$ is the distance scaling factor, as a suboptimal alternative for cases where direct communication is not possible.

Denote by $ T^U_{ks}$ the time duration taken for communication of SOV $s$ in the $k$-th communication round, which is expressed as
\begin{equation}
    T^U_{ks}=\sum_{m \in \mathcal{M}} \sum_{r \in \mathcal{H} \setminus s}\alpha^m_{ks}u^m_{kr} \frac{Z^m}{R_{ksr}},
    \label{eq:time-com-SOV}
\end{equation}
where $Z^m$ represents the size of the model parameter $\boldsymbol{w}^m$.
Please note the definitions of $\alpha^m_{ks}$ and $u^m_{kr}$, with $\alpha^m_{ks}u^m_{kr} \in \{ 0,1 \}$.
The product $\alpha^m_{ks}u^m_{kr}$ represent the communication between the CHV and the corresponding SOVs.
Denote by $E^U_{ks}$ the energy consumption due to communication transmission of SOV $s$ in the $k$-th communication round, which is expressed as
\begin{equation}
    E^U_{ks}= \sum_{m \in \mathcal{M}} \sum_{r \in \mathcal{H} \setminus s}\alpha^m_{ks}u^m_{kr} p_{ks}\frac{Z^m}{R_{ksr}}.
    \label{eq:energy-com-SOV}
\end{equation}

The CHV and the corresponding SOVs communicate in parallel on different subcarriers. 
Denote by $T^U_{kr}$ the communication time duration required by CHV $r$ in the $k$-th communication round, which is expressed as
\begin{equation}
    T^U_{kr}= \sum_{m \in \mathcal{M}} u^m_{kr} \max_{s\in\mathcal{H}}\{(\alpha^m_{ks}-u^m_{ks})T^U_{ks}\}.
    \label{eq:time-com-CHV}
\end{equation}
Denote by $E^U_{kr}$ the communication energy consumption of CHV $r$ in the $k$-th communication round, which is expressed as
\begin{equation}
    E^U_{kr}= \sum_{m \in \mathcal{M}} u^m_{kr} \left( \sum_{s\in\mathcal{H}}(\alpha^m_{ks}-u^m_{ks})E^U_{ks} \right).
    \label{eq:energy-com-CHV}
\end{equation}

\subsection{Computation Model}

Denote by $p^C_{kh}$ the local computing power of vehicle $h$ in the $k$-th communication round. Based on the definition in \cite{zhang2024vehicle}, $p^C_{kh}$ is expressed as
\begin{equation}
    p^C_{kh}=\lambda f_{kh}^3,
    \label{eq:computing-power}
\end{equation}
where $\lambda$ represents the effective switching capacitance of the CPU, and $f_{kh}$ represents the CPU frequency of vehicle $h$ in the $k$-th communication round.
In the $k$-th communication round, vehicle $h \in \mathcal{H}$ needs to train the model for the assigned task $m$.
Denote by $T^C_{kh}$ the computation time required by vehicle $h$, which is expressed as
\begin{equation}
    T^C_{kh}=\sum_{m \in \mathcal{M}} \alpha^m_{kh} \frac{I |\mathcal{D}^m_h| q}{f_{kh}},
    \label{eq:time-cmp}
\end{equation}
where $q$ represents the CPU frequency required to process 1 bit.
The computation energy consumption of vehicle $h$ in the $k$-th communication is expressed as
\begin{equation}
    E^C_{kh}=p^C_{kh}T^C_{kh}=\sum_{m \in \mathcal{M}} \alpha^m_{kh} I \lambda |\mathcal{D}^m_h| y f_{kh}^2.
    \label{eq:energy-cmp}
\end{equation}

\subsection{TSLP Model}

\begin{figure}[H]
  \centering
  \includegraphics[width=0.48\textwidth]{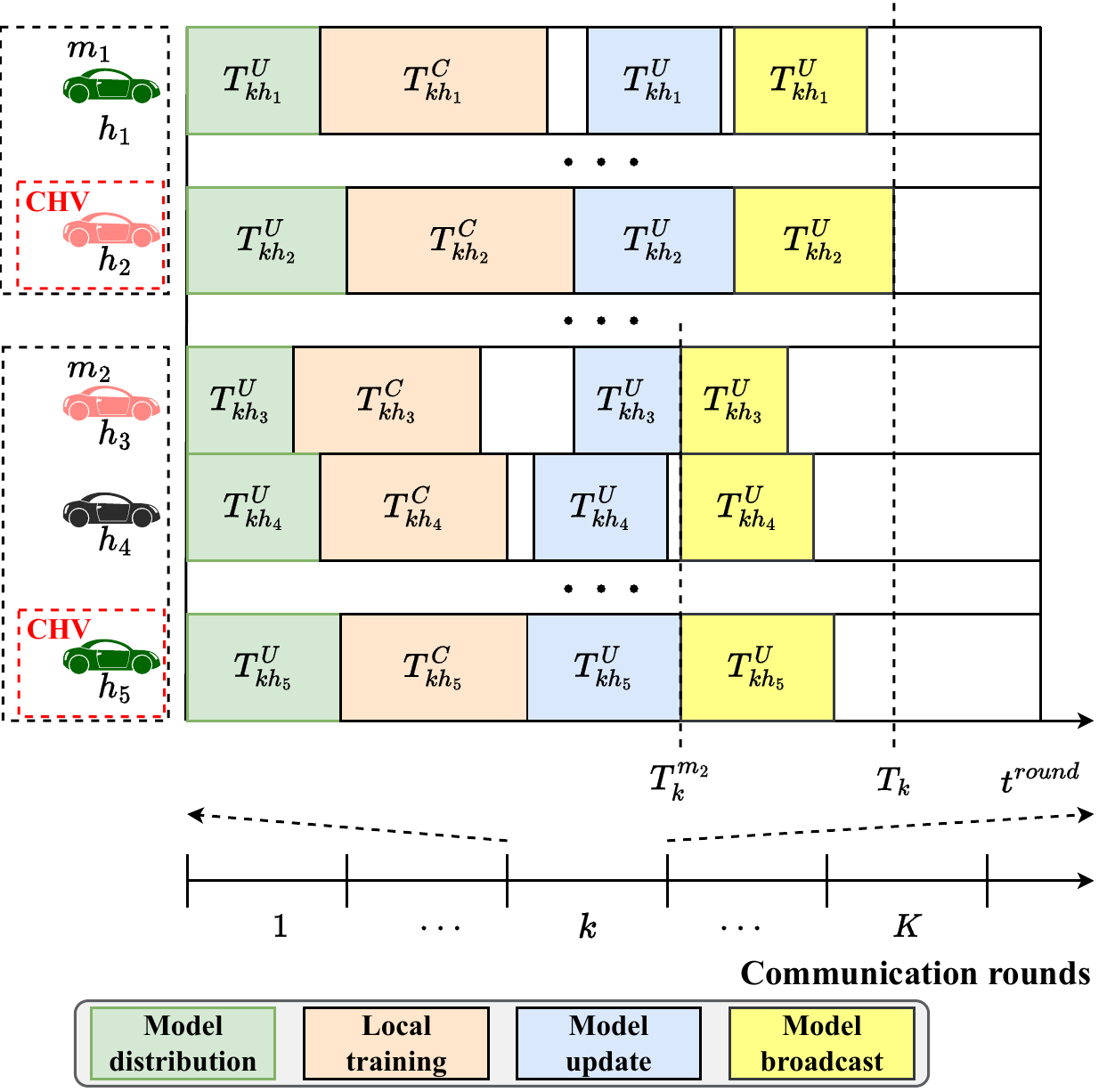}\\
  \caption{Timeline.}
  \label{fig:timeline}
\end{figure}

The timeline of MMFL is shown in Fig. \ref{fig:timeline}.
In each communication round, there is a synchronization point after the model aggregation process.
Before the synchronization point, vehicle $h$ undergoes model distribution, model update, and local training.
The communication time duration for both model distribution and model update is $T^{U}_{kh}$.
The computing time duration of model update is $T^{C}_{kh}$.
In addition, the aggregation process is not incorporated into our timeline illustration due to its comparatively short duration.
Thus, denote by $T^m_k$ the synchronization point of task $m$ in the $k$-th communication round, which is expressed as
\begin{equation}
    T^m_{k}=\max_{h}\{\alpha^m_{kh}(T^C_{kh}+2T^{U}_{kh})\}.
    \label{eq:time-sum-m}
\end{equation}
After the synchronization point, vehicle $h$ undergoes model broadcast.
The communication time duration of model broadcast is $T^{C}_{kh}$.
Thus, denote by $T_k$ the total time consumption in the $k$-th communication round, which is expressed as
\begin{equation}
    T_k=\max_{h}\{\sum_{m \in \mathcal{M}}\alpha^m_{kh}T^m_{k}+T^{U}_{kh}\}.
    \label{eq:time-sum}
\end{equation}

In a single communication round, vehicle $h$ needs to go through three communication processes, namely model distribution, model update, and model broadcast.
The communication energy consumption of each process is $E^{U}_{kh}$.
In addition, the vehicle also needs to undergo a local training process, where the computing energy consumption is $E^C_{kh}$.
Thus, denote by $E_{kh}$ the total energy consumption of vehicle $h$ in the $k$-th communication round, which is expressed as
\begin{equation}
    E_{kh}=E^C_{kh}+3E^{U}_{kh}.
    \label{eq:energy-sum}
\end{equation}

Denote by $E^{res}_{kh}$ the residual energy of vehicle $h$ in the $k$-th communication round, which is expressed as
\begin{equation}
    E^{res}_{kh}=E_h-\sum_{j=1}^{k-1}E_{jh},
    \label{eq:energy-res}
\end{equation}
where $E_h$ represents the initial energy of vehicle $h$.

Denote by $F^{loss}$ the average loss of each task over the K-th (i.e., the last) communication round, which is expressed as
\begin{equation}
    F^{loss}=
        \frac{1}{M}
        \sum_{m\in\mathcal{M}}
        \left (
        \frac{1}{{\sum}_{h\in\mathcal{H}}\alpha^m_{Kh}}
        \sum_{h\in\mathcal{H}}
        \alpha^m_{Kh}f^m_h(\boldsymbol{w}^m_{Kh})
        \right ).
    \label{eq:F-loss}
\end{equation}
In this paper, we aim to minimize the loss function for each training task subject to latency and energy consumption.
Denote by $\rho^m_{kh}$ the communication round index that corresponds to the most recent participation of vehicle $h$ for task $m$ prior to the $k$-th communication round.
The joint optimization problem
for task scheduling, communication bandwidth allocation, and leader selection (TSLP) is given in (\ref{eq:final-target0}).
\begin{figure}[htbp]
\begin{subequations}
\begin{alignat}{2}
& \mathcal{P}0: \min _{\boldsymbol{\alpha},\boldsymbol{u},\boldsymbol{c}} F^{loss} \label{a0} \\
\text{s.t}. \quad
& \max _{k}\left( \alpha^m_{kh} T_{k}\right) \leq t^{round},h\in\mathcal{H},m\in\mathcal{M},k\in\mathcal{K}, \label{b0}\\
& \alpha^m_{kh} E_{kh}\leq E^{res}_{kh},h\in\mathcal{H},m\in\mathcal{M},k\in\mathcal{K}, \label{c0}\\
& \sum_{m\in\mathcal{M}}\alpha^m_{kh} \leq 1,h\in\mathcal{H}, k \in \mathcal{K},\label{d0} \\
& u^m_{kh} \leq \alpha^m_{kh}, m\in\mathcal{M}, k\in\mathcal{K}, h\in\mathcal{H},\label{e0} \\
&\sum_{h\in\mathcal{H}}\alpha^m_{kh}u^m_{kh}=1,m\in\mathcal{M}, k \in \mathcal{K},\label{f0} \\
&\sum_{h \in \mathcal{H}} c^m_{kvn} \leq 1, n \le N, m \in \mathcal{M}, k \in \mathcal{K}, \label{g0} \\
&  u^m_{kr}=\arg \max \limits_{\boldsymbol{u}^m_k} \rho^m_{kh}, m \in \mathcal{M}, k \in \mathcal{K}, \label{h0} \\
&  {\alpha^m_{kh} \in \lbrace 0,1\rbrace },  {u^m_{kh} \in \lbrace 0,1\rbrace }, {c^m_{kvn} \in \lbrace 0,1\rbrace } \label{i0}.
\end{alignat}
\label{eq:final-target0}
\end{subequations}
\end{figure}

In (\ref{eq:final-target0}), constraint (\ref{b0}) ensures that the time spent by each vehicle in any communication round is strictly bounded by a time duration of $t^{round}$.
The constraint is influenced by $\boldsymbol{\alpha}$, $\boldsymbol{u}$, and $\boldsymbol{c}$, as seen in (\ref{eq:bandwidth-allocation}), (\ref{eq:transmission-rate}), (\ref{eq:time-com-SOV}), (\ref{eq:time-com-CHV}), (\ref{eq:time-sum-m}) and (\ref{eq:time-sum}).
This implies that the variables are coupled.
Constraint (\ref{c0}) ensures that the energy consumption of each participating vehicle in any communication round does not exceed its current energy capacity.
Constraints (\ref{d0}) and (\ref{e0}) state that a vehicle participates in at most one task.
Constraint (\ref{f0}) represents that each task is allocated a cluster head.
Constraint (\ref{g0}) indicates orthogonal allocation of subcarriers.
Constraint (\ref{h0}) is used to mitigate the issue that a model update becomes overwritten.
Constraint (\ref{i0}) sets the range of values that the optimization variable can take.

\section{Problem Analysis}

In this section, we first discuss the convergence bounds for a single task under the MMFL framework.
For multi-task scenarios, we model TSLP as a resource allocation game, and then prove the existence of NE for the game.

\subsection{Convergence Analysis}

For each task $m$, we make the following assumptions \cite{yan2024dynamic}:

\textit{Assumption 1:} The local loss function $f^m_{h}(\boldsymbol{w}^m)$ is $L$-smooth for each vehicle $h \in \mathcal{H}$ in each communication round $k \in \mathcal{K}$, i.e,
\begin{equation}
    \begin{aligned}
        & f^m_{h}(\boldsymbol{w}^m_{k}) - f^m_{h}(\boldsymbol{w}^m_{k-1}) \\
        & \leq\left\langle\nabla f^m_{h}(\boldsymbol{w}^m_{k-1}),\boldsymbol{w}^m_k-\boldsymbol{w}^m_{k-1}\right\rangle+\frac{L}{2}\left\|\boldsymbol{w}^m_k-\boldsymbol{w}^m_{k-1}\right\|^2.
    \end{aligned}
    \nonumber
\end{equation}

\textit{Assumption 2:} The local loss function $f^m_{h}(\boldsymbol{w}^m)$ is $\mu$-strongly convex for each vehicle $h \in \mathcal{H}$ in each communication round $k \in \mathcal{K}$, i.e,
\begin{equation}
    \begin{aligned}
        & f^m_{h}(\boldsymbol{w}^m_{k}) - f^m_{h}(\boldsymbol{w}^m_{k-1}) \\
        & \geq\left\langle\nabla f^m_{h}(\boldsymbol{w}^m_{k-1}),\boldsymbol{w}^m_k-\boldsymbol{w}^m_{k-1}\right\rangle+\frac{\mu}{2}\left\|\boldsymbol{w}^m_k-\boldsymbol{w}^m_{k-1}\right\|^2.
    \end{aligned}
    \nonumber
\end{equation}

\textit{Assumption 3:} The stochastic gradient is unbiased, i.e.,
\begin{equation}
    \underset{\boldsymbol{x}\sim\mathcal{D}^m_{h}}{\operatorname*{\mathbb{E}}}
    \left[\nabla f^m(\boldsymbol{w}^m;\boldsymbol{x})\right]=\underset{h\sim\mathcal{P}_{h}}{\operatorname*{\mathbb{E}}}\left[\nabla f^m_{h}(\boldsymbol{w}^m)\right]=\nabla F^m(\boldsymbol{w}^m).
    \nonumber
\end{equation}

\textit{Assumption 4:} The stochastic gradient is variance-bounded, i.e.,
\begin{equation}
    \underset{x\sim\mathcal{D}^m_{h}}{\operatorname*{\mathbb{E}}}\left[\left\|\nabla f^m(\boldsymbol{w}^m;\boldsymbol{x})-\nabla F^m(\boldsymbol{w}^m)\right\|^2\right]\leq G^2.
    \nonumber
\end{equation}

\textit{Assumption 5:} In the $k$-th communication round beyond the initial one, the CHV of task $m$ is selected from the cluster of task $m$ in the $(k-1)$-th communication round.

Assumption 5 prevents outdated models from overwriting new updates.
Based on the above assumptions, we obtain the following lemma:

\textbf{Lemma 1.} \textit{With Assumptions 1-5 and the aggregation rule (\ref{eq:aggregation-define}), the expected decrease of loss after one round is upper bounded by}
\begin{equation}
    \begin{aligned}
        & \mathbb{E} \left[ F^m(\boldsymbol{w}^m_k) \right] - \mathbb{E} \left[ F^m(\boldsymbol{w}^m_{k-1}) \right] \leq \\
        & \eta_k \left( \frac{L\eta_k}{2} - 1 \right) \left\|\nabla F^m(\boldsymbol{w}^m_{k-1}) \right\|^2 + \frac{L\eta_k^2}{2} \frac{G^2}{B_{k}\sum_{h\in\mathcal{H}}\alpha^m_{kh}},
    \end{aligned}
    \label{eq:lemma-1}
\end{equation}
\textit{where the expectation is taken over the randomness of SGD.}

\begin{proof}
    See Appendix \ref{appendix-a}.
\end{proof}

Based on Lemma 1, we obtain the convergence performance after $K$ communication rounds, which is presented in Theorem 1.

\textbf{Theorem 1.} \textit{After $K$ communication rounds of training, the difference between $F^m(\boldsymbol{w}^m_K)$ and the optimal global loss function denoted by $F^m(\boldsymbol{w}^{m*})$ is upper bounded by}
\begin{equation}
    \begin{aligned}
        & \mathbb{E} \left[ F^m(\boldsymbol{w}^m_K) \right] - F^{m}(\boldsymbol{w}^{m*}) \\
        & \leq \left( \mathbb{E} [ F^m(\boldsymbol{w}^m_{0}) ] -F^m(\boldsymbol{w}^{m*}) \right) \prod_{k=1}^{K}(1-\mu\eta_{k}) \\
        & + \sum_{k=1}^{K-1}\frac{\eta_k}{2}\frac{G^2}{B_{k}\sum_{h\in\mathcal{H}}\alpha^m_{kh}} \prod_{j=k+1}^{K}(1-\mu\eta_{j}) \\
        & + \frac{\eta_K}{2}\frac{G^2}{B_{K}\sum_{h\in\mathcal{H}}\alpha^m_{Kh}}.
    \end{aligned}
    \label{eq:theorem-1}
\end{equation}

\begin{proof}
    See Appendix \ref{appendix-b}.
\end{proof}

According to Theorem 1, we have the following insights.
First, the convergence bound of the global loss function for task $m$ decreases with the increase of the number of participants, indicating that the decision of $\boldsymbol{\alpha}$ directly affects the convergence of an individual FL task.
Second, the variable $\boldsymbol{u}$ affects the update status of an individual FL task, i.e., the extent to which model updates are overwritten.
It also influences the resource consumption of each vehicle due to mobility.
Furthermore, optimizing the bandwidth allocation variable $\boldsymbol{c}$ helps to reduce the operational latency of the framework.
Given that $\mu > 0$ and $\eta_k > 0$, we have $1 - \mu\eta_k < 1$, which indicates that as the number of communication rounds increases, the convergence bound of the global loss function for task $m$ decreases.

\subsection{Resource Allocation Game Formulation}
\label{sec:game-model}

For multiple task scenarios, the convergence rate of the loss  within the framework's operation is influenced by $\boldsymbol{\alpha}$, $\boldsymbol{u}$, and $\boldsymbol{c}$, for which the relations are not analytically available.
By Theorem 1, we find that the convergence bound of a single FL task is directly related to the number of participants.
Note that the loss function represents the distance between the predicted labels and the true labels, which are used to assess the model's fitness during training.
This value decreases as the training progresses.
Accuracy represents the agreement between the predicted labels and the true labels, and is used to measure the model's performance on the test set.
The closer this value is to one, the better the model's performance.
The work in \cite{li2021efficient} models the relationship between accuracy and training efficiency in multi-task FL.
The transition between minimizing the loss function and maximizing training efficiency does not affect the objective of TSLP.

Consequently, we adopt the aggregate training efficiency as a measure. 
A higher training efficiency implies that the number of communication rounds $K$ required for convergence is smaller, thereby enhancing energy efficiency. 
The training efficiency is defined in accordance with the work in \cite{li2021efficient}.

\textbf{Definition 1.} The training efficiency of task $m$ in the $k$-th communication round increases with the number of assigned vehicles, but the increase in training efficiency is diminishing, roughly following a logarithmic relationship, which is expressed as
\begin{equation}
    \psi^m_k=\beta^m \log(\sum_{h \in \mathcal{H}}\alpha^m_{kh})+\theta^m.
    \label{eq:def-1}
\end{equation}

We model the problem as a non-cooperative game among the tasks, where each task is regarded as a player and it independently decides on its vehicle allocation strategy, with the goal of faithfully improving its own training efficiency. The TSLP in (\ref{eq:final-target0}) is transformed into
\begin{figure}[htbp]
\begin{subequations}
\begin{alignat}{2}
& \mathcal{P}1: \max _{\boldsymbol{\alpha},\boldsymbol{u},\boldsymbol{c}} \sum_{k \in \mathcal{K}} \sum_{m \in \mathcal{M}} \psi^m_k \label{a1} \\
\text{s.t}. \quad
& (\ref{b0})-(\ref{i0}). \label{b1}
\notag
\end{alignat}
\label{eq:final-target1}
\end{subequations}
\end{figure}

To describe the competitive relationship between tasks in the $k$-th communication round, we define resource allocation game ${\mathcal{G}_k}=\left\{\mathcal{M},\mathbb{S}_k,\{U^m_k\}_{m \in \mathcal{M}}\right\}$ \cite{Xu2023Joint}, where $\mathbb{S}_k$ represents the strategy space of the game, defined as the Cartesian product of all individual strategy sets of the tasks: $\mathbb{S}_k= \mathbf{S}^1_k\times\cdots\times\mathbf{S}^m_k\times\cdots\times\mathbf{S}^M_k$, where $\mathbf{S}^m_k$ represents the set of all strategies for task $m$ in the $k$-th communication round.
Each $\mathscr{S}_k\in\mathbb{S}_k$ is a strategy profile.
For task $m$, the profile $\mathscr{S}_k=(\mathscr{S}^1_k,...,\mathscr{S}^m_k,...,\mathscr{S}^M_k)$ can be rewritten as $\mathscr{S}_k=(\mathscr{S}^m_k,\mathscr{S}^{-m}_k)$, where $\mathscr{S}^m_k$ denotes the strategy of task $m$ in the $k$-th communication round,
represented as $\mathscr{S}^m_k=\{ h \in \mathcal{H} \mid \alpha^m_{kh}=1\}$,
and $\mathscr{S}^{-m}_k$ represents the joint strategy adopted by the tasks other than task $m$. Denote by $U^m_k(\mathscr{S}_k)$ the utility function of task $m$ in the $k$-th communication round, which is defined as follows.

\textbf{Definition 2.} The utility function $U^m_k(\mathscr{S}_k):\mathbb{S}_k\mapsto\mathbb{R}$ is defined as the training efficiency of task $m$ under the strategy configuration $\mathscr{S}_k$ , where $\mathbb{R}$ represents the set of real numbers. Specifically, $U^m_k(\mathscr{S}_k)$ is expressed as
\begin{equation}
    \begin{aligned}
        U^m_k(\mathscr{S}_k)=
        \left\lbrace\begin{array}{ll} 
        \psi^m_k , & \text{if } \text{(\ref{b0})-(\ref{i0}) are satisfied,} \\
        0, & \text{otherwise.} 
        \end{array}
        \right.
    \end{aligned}
    \label{eq:def-2}
\end{equation}

Next, we demonstrate that the game model $\mathcal{G}_k$ is a potential game (PG) with at least one NE by providing the potential function. We first present the definition of a potential game.

\textbf{Definition 3.} If there exists a potential function $\Omega^m_k(\mathscr{S}_k)$ that satisfies (\ref{eq:def-3}), then the game is a potential game.
\begin{equation}
    \begin{aligned}
        & U^m_k (\mathscr{S}^{m{\prime}}_k,\mathscr{S}^{-m}_k) > U^m_k(\mathscr{S}^m_k,\mathscr{S}^{-m}_k) \\
        & \mapsto
        \Omega^m_k(\mathscr{S}^{m{\prime}}_k,\mathscr{S}^{-m}_k) >
        \Omega^m_k(\mathscr{S}^m_k,\mathscr{S}^{-m}_k).
    \end{aligned}
    \label{eq:def-3}
\end{equation}

\textbf{Theorem 2.} \textit{For the potential function defined below for task $m$, game $\mathcal{G}_k$ is a potential game.}
\begin{equation}
    \Omega^m_k(\mathscr{S}_k)=\sum_{i \in \mathcal{M}} \left[ U^i_k(\mathscr{S}^m_k,\mathscr{S}^{-m}_k)-U^i_k(-\mathscr{S}^m_k,\mathscr{S}^{-m}_k) \right],
    \label{eq:theorem-2}
\end{equation}
\textit{where $U^i_k(-\mathscr{S}^m_k,\mathscr{S}^{-m}_k)$ represents the utility that can be achieved when $\mathscr{S}^m_k$ is ineffective, meaning that no valid set of $\boldsymbol{\alpha}$, $\boldsymbol{u}$ and $\boldsymbol{c}$ can satisfy this strategy.}

\begin{proof}
    See Appendix \ref{appendix-c}.
\end{proof}

In the game model $\mathcal{G}_k$, all tasks attempt to achieve a NE by maximizing their utility in the presence of conflicting interests. Since Theorem 2 proves that the resource allocation game is a potential game, it has the finite improvement property (FIP), thus a NE allocation strategy can be obtained through a finite number of iterations \cite{Chen2024qoe}. Solving the NE in the resource allocation game can enhance the training efficiency of all tasks in the framework, improving the convergence rate.

\section{Algorithm Design}
Based on the proof of the NE existence, we first reformulate the problem in (\ref{eq:final-target1}) as a DEC-POMDP, and then propose an HAPPO-based algorithm to solve DEC-POMDP.

\subsection{DEC-POMDP Reformulation} \label{DEC-POMDP}

\begin{figure}[!htbp]
  \centering
  \includegraphics[width=0.48\textwidth]{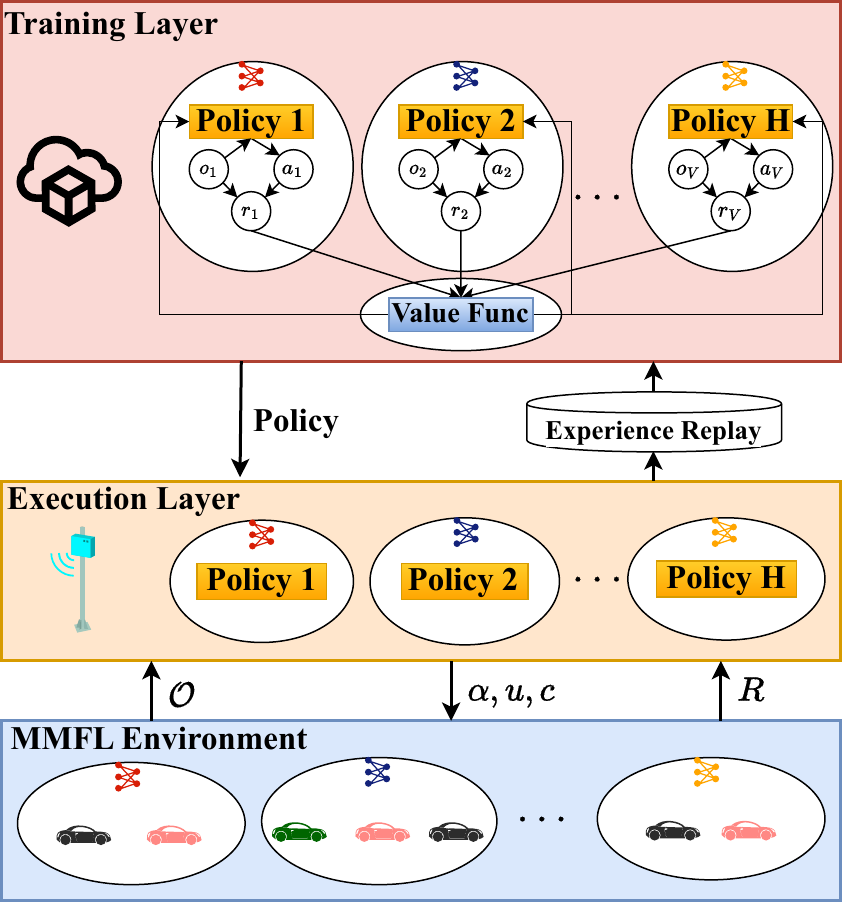}\\
  \caption{Diagram of centralized training and distributed execution framework.}
  \label{fig:agents}
\end{figure}

As shown in Fig. \ref{fig:agents}, we design a centralized training and decentralized execution multi-agent reinforcement learning (MARL) algorithm to solve the resource allocation game.
After the vehicles submit training requests, the cloud server creates agents corresponding to the vehicles and sends the policies to the nearest available RSU to guide the FL process.
In each communication round, each CHV collects the observation of the vehicles in its cluster, forms its reward, and sends them to the RSU.
The RSU stores the records and submits them to the cloud server as a replay buffer.
The cloud server then extracts several batches from the replay buffer for training.

Denote by $\mathcal{\zeta}=
\Bigl \lbrace \{ \mathcal{\zeta}_h \}_{h=1}^H \Bigr \rbrace$ the agent set of task distributors, where each agent corresponds to a vehicle.
The resource allocation game (\ref{eq:final-target1}) is reformulated to a DEC-POMDP ${\langle}\mathcal{S},\mathcal{A},\mathcal{O},R,P\rangle$, where $\mathcal{S}$, $\mathcal{A}$, $\mathcal{O}$, $R$, $P$ represent the global state, action set, local observation set, reward function, and state transition function, respectively \cite{yu2024Multi}.
The local observation space, denoted by $o_{kh}$, consists of the most recent communication round index of the vehicle's participation in any task, remaining energy, transmission rate, position, and speed of vehicle $h$ in the $k$-th communication round, which is expressed as
    \begin{equation}
    \begin{aligned}
        o_{kh}=
    \Bigl \lbrace 
    \{ \rho^m_{kh} \}_{m \in \mathcal{M}}, E^{res}_{kh}, \bigr . 
    \bigl . \sum_{r \in \mathcal{H}}\alpha^m_{kh} u^m_{kr} R_{khr},x_{kh},y_{kh},s_{kh}
    \Bigr \rbrace.
    \end{aligned}
    \label{eq:observation-task}
    \end{equation}

The global state $\mathcal{S}$ contains the local observations of all agents across all $K$ communication rounds, which is expressed as
\begin{equation}
\begin{aligned}
    \mathcal{S}=
    \Bigl \lbrace 
    o_{kh} | h \in \mathcal{H}, k \in \mathcal{K}
    \Bigr \rbrace.
\end{aligned}
    \label{eq:global-state}
\end{equation}

For the decision of $\boldsymbol{\alpha}_k$, denote by $a_v$ the action space of the agent corresponding to vehicle $h$. In each communication round, vehicle $h$ can choose any task, corresponding to the task indexes in the action space; it can also choose not to participate in any task, corresponding to value zero. The action space is expressed as
\begin{equation}
        a_v=\{ 0,1,2,...,M \}.
        \label{eq:action-task}
\end{equation}
For the decision of $\boldsymbol{u}_k$ and $\boldsymbol{c}_k$, we need to ensure a balance between the efficient execution of FL and resource consumption.
Specifically, we propose a joint optimization for leader selection and subcarrier allocation, as shown in Algorithm \ref{algo:leader-selection-bandwidth-allocation}.
We score all vehicles within the task group $m$. In Line (\ref{eq:score}), the first term considers communication efficiency and the second term considers training efficiency, where $\varepsilon$ represents the weighting coefficient.
Denote by $h^{m*}_k$ the vehicle with the highest score in the group, which is then selected as the leader for task $m$ in the $k$-th communication round.
Note that $\boldsymbol{\alpha}_k$ is fixed at this point.
The complexity of Algorithm \ref{algo:leader-selection-bandwidth-allocation} is $O(M \cdot N \log H)$.
By applying this algorithm, the dimensionality of the action space is effectively reduced, leading to improved search efficiency.
\begin{algorithm}[!ht]
    \renewcommand{\algorithmicrequire}{\textbf{Input:}}
	\renewcommand{\algorithmicensure}{\textbf{Output:}}
	\caption{Joint optimization of leader selection and subcarrier allocation}
    \label{algo:leader-selection-bandwidth-allocation}
    \begin{algorithmic}[1] 
        \REQUIRE  $\boldsymbol{\alpha}_k$, $\{\rho^m_{kh} | m \in \mathcal{M}, h \in \mathcal{H}\}$, and $\{(x_{kh}, y_{kh}) | h \in \mathcal{H}\}$; 
	    \ENSURE $\boldsymbol{u}_k$ and $\boldsymbol{c}_k$; 
        \STATE // At the beginning of the $k$-th communication round;
        \STATE \textbf{Initialize:} $\boldsymbol{u}_k \gets \mathbf{0}, \, \boldsymbol{c}_k \gets \mathbf{0}$;
        \FORALL {$m \in \mathcal{M}$}
            \FORALL {$r \in \mathcal{H} \textbf{ with } \alpha^m_{kr} = 1$}
                \STATE \textbf{Initialize:} Max-Heap $Q = \emptyset$;
                \FORALL {$h \in \mathcal{H} \setminus r$ \textbf{ with } $\alpha^m_{kh} = 1$}
                    \STATE Assign a subcarrier to vehicle $h$;
                    \STATE Calculate $T^U_{kh}$ based on (\ref{eq:time-com-SOV}) and (\ref{eq:time-com-CHV});
                    \STATE Insert vehicle $h$ with key $T^U_{kh}$ into $Q$;
                \ENDFOR
                \WHILE {Remaining subcarriers $>$ $0$}
                    \STATE $h \gets$ Pop the top element from $Q$;
                    \STATE Assign a subcarrier to vehicle $h$;
                    \STATE Calculate $T^U_{kh}$ based on (\ref{eq:time-com-SOV}) and (\ref{eq:time-com-CHV});
                    \STATE Insert vehicle $h$ with key $T^U_{kh}$ into $Q$;
                \ENDWHILE
                \STATE $\hat{h} \gets$ Pop the top element from $Q$;
                \STATE Calculate $T^U_{k\hat{h}}$ based on (\ref{eq:time-com-SOV}) and (\ref{eq:time-com-CHV});
                \STATE Calculate the score for vehicle $r$: 
                \label{eq:score}
                \begin{equation}
                    score_r =  \frac{\rho^m_{kr}}{k} - \varepsilon \frac{T^U_{k\hat{h}}}{t^{round}} ; 
                \end{equation}
            \ENDFOR
            \STATE $h^{m*}_k = \arg\max_{h \in \mathcal{H}} score_v$;
            \STATE $u^m_{k,h^{m*}_k} = 1$;
            \STATE Generate $\boldsymbol{c}^m_{k}$ for $h^{m*}_k$ according to the above process;
        \ENDFOR
    \end{algorithmic}
\end{algorithm}

Denote by $r_{kh}$ the local reward of the agent corresponding to vehicle $h$ in the $k$-th communication round; the reward includes the utility of the corresponding task in game (\ref{eq:final-target1}).
Additionally, we introduce $\rho^m_{kh}$ with the aim of encouraging the agent to satisfy Assumption 5 as much as possible. The local reward is expressed as
\begin{equation}
    r_{kh}=\sum_{m \in \mathcal{M}} \alpha^m_{kh} \psi^m_k \rho^m_{kh}.
    \label{eq:reward-local}
\end{equation}
Furthermore, for action outputs that do not satisfy the constraints of game (\ref{eq:final-target1}), we apply a penalty to all agents.

The global reward is defined as the sum of the local rewards (\ref{eq:reward-local}) over all $K$ communication rounds and all agents in the framework. The global reward is expressed as
\begin{equation}
    R=\sum_{k \in \mathcal{K}} \sum_{h \in \mathcal{H}} r_{kh}.
    \label{eq:reward-global}
\end{equation}

For the transition of local observations $P$ from $o_{kh}$ to $o_{k+1,h}$, the remaining energy is updated by (\ref{eq:energy-res}), the position and speed of the vehicle are acquired in real time, and the transmission rate is estimated based on the current conditions and (\ref{eq:transmission-rate}).

\subsection{HAPPO-based Optimization Algorithm}

Denote by $\pi_k$ the joint decision made by $\mathcal{\zeta}$ in the $k$-th communication round, and $(\pi_k)_{k=0}^{\infty}$ the sequence of joint decisions.
Sequence $(\pi_k)_{k=0}^{\infty}$ obtained through the HAPPO algorithm possesses four properties.
First, the expected return of the joint policy is monotonically increasing.
Second, as the number of iterations increases, the generated value functions converge to the value function corresponding to a NE.
Third, the expected return of the joint policy converges to the expected return of a NE.
Finally, the $\omega$-limit set of the policy sequence contains NE strategies \cite{zhong2024heterogeneous}.
Furthermore, using HAPPO allows for a clearer description of the interactions between agents, and it converges faster compared to traditional reinforcement learning (RL).

To solve DEC-POMDP in Section \ref{DEC-POMDP} and obtain a NE solution, we propose an HAPPO-based optimization algorithm.
The goal of the proposed algorithm is to train $H$ local policy networks and one global value network.

Denote by $\pi_{\theta^{h}}$ the policy of agent $\zeta_h$, which is formulated by a multilayer perception (MLP) neural network with parameter $\theta^{h}$.
Denote by $V_{\phi}$ the global V-value function, which is formulated by an MLP neural network with parameter $\phi$.
Here, $\pi_{\theta^{h}}$ is a multinomial distribution used for action selection, i.e.,
\begin{equation}
    a^h \sim \text{Categorical}(\pi_{\theta^h}(a^h \mid o^h)),
    \label{eq:sampling}
\end{equation}
where $a^h$ represents the action taken by agent $\zeta_h$ in the local observation $o^h$.
We use $V_{\phi}$ to represent the expected reward for the entire system when all agents act according to their respective policies $\pi_{\theta^{h}}$, given the global state $s$, i.e.,
\begin{equation}
    V_\phi(s) = \mathbb{E}_{\pi_{\theta^1}, \pi_{\theta^2}, \dots, \pi_{\theta^H}} \left[ \sum_{t=0}^{\infty} \gamma^t R_t \mid s \right], \nonumber
\end{equation}
where $R_t$ represents the joint reward of all agents at time step $t$, and $\gamma$ represents the discount factor.
Denote by $T_s$ the timesteps for trajectory collection. At time step $t$, the expect reward $\hat{R}_t$ is expressed as
\begin{equation}
    \hat{R}_t = \sum_{m=0}^{T_s - t - 1} \gamma^m R_m.
    \label{eq:expect-reward}
\end{equation}
Note that one time step in the algorithm refers to executing one FL communication round in the MMFL environment. When we reset the MMFL environment, the FL communication round is set to zero, but the time step remains unchanged.

The advantage function in reinforcement learning is used to estimate the advantage of a particular action relative to the average policy.
Denote by $\hat{A}_t(s,\boldsymbol{a})$ the advantage function at time step $t$. It is calculated by generalized advantage estimation (GAE), which is expressed as
\begin{equation}
    \hat{A}_t(s,\boldsymbol{a}) = \sum_{m=0}^{T_s - t - 1} (\gamma\beta)^{m}\delta_{t+m},
    \label{eq:advantage-function}
\end{equation}
where $\boldsymbol{a}$ represents the joint action, $\beta$ represents the smoothing factor, and $\delta_t$ represents the temporal difference error, which is expressed as
\begin{equation}
    \delta_t=R_t+\gamma V_\phi(s_{t+1})-V_\phi(s_t). \nonumber
\end{equation}

Denoting by $\epsilon$ the discount ratio, the training of the proposed HAPPO-based optimization algorithm is shown in Algorithm \ref{algo:happo}. Note that the main difference from the previous MAPPO algorithm \cite{Yu2021MAPPO} lies in Lines \ref{sequence-update-start} to \ref{sequence-update-end}. In HAPPO, the policy networks of individual agents are updated sequentially, with no shared policy parameters between agents \cite{kuba2021trust}. This approach mitigates conflicts in policy updates among agents, thereby improving the training efficiency.
\begin{algorithm}[!ht]
    \renewcommand{\algorithmicrequire}{\textbf{Input:}}
	\renewcommand{\algorithmicensure}{\textbf{Output:}}
	\caption{Training of HAPPO-based optimization algorithm}
    \label{algo:happo}
    \begin{algorithmic}[1] 
        \REQUIRE  Batch size $B$, number of agents $H$, episodes $K_s$, steps per episode $T_s$;
        \ENSURE Actor networks $\{\theta^h_{K_s} | \zeta_h \in \zeta \}$, global V-value network $\phi_{K_s}$;
	    \STATE \textbf{Initialize:} Actor networks $\{\theta^h_0 | \zeta_h \in \zeta \}$, global V-value network $\phi_0$, replay buffer $\mathcal{B}$;
        \FOR {$k_s = 0,1,\dots,K_s-1$}
            \STATE Collect a set of trajectories by running the joint policy $\boldsymbol{\pi}_{\boldsymbol{{\theta}}_{k_s}} = (\pi^1_{\theta^1_{k_s}},\dots,\pi^H_{\theta^H_{k_s}})$;
            \FOR{$t = 0,1,\dots,T_s-1$}
                \STATE Obtain $\boldsymbol{\alpha}_t$ based on sampling (\ref{eq:sampling});
                \STATE Obtain $\boldsymbol{u}_t, \boldsymbol{c}_t$ based on Algorithm \ref{algo:leader-selection-bandwidth-allocation};
                \STATE Execute a communication round $t$ of the MMFL environment using $\boldsymbol{\alpha}_t, \boldsymbol{u}_t, \boldsymbol{c}_t$;
                \IF{Constraints (\ref{b0})-(\ref{i0}) is not all satisfied}
                    \STATE Reset the MMFL environment;
                \ENDIF
                \STATE Obtain $\{ (o_{th},r_{th}) \mid \forall \zeta_h \in \zeta \}$;
            \ENDFOR
            \STATE Push transitions $\{(o_{th},a_{th},o_{t+1,h},r_{th}) \mid \forall \zeta_h \in \zeta, t \leq T_s \}$ into $\mathcal{B}$;
            \STATE Sample a random minibatch of $B$ transitions from $\mathcal{B}$;
            \STATE Compute advantage function $\hat{A}(s,\boldsymbol{a})$ based on (\ref{eq:advantage-function});
            \STATE Draw a random permutation of agents $\zeta_{1:H}$; \label{sequence-update-start}
            \STATE Set $M^{\zeta_1}(s, \boldsymbol{a}) = \hat{A}(s,\boldsymbol{a})$;
            \FORALL{$\zeta_h \in \zeta$}
                \STATE Update actor network $\theta^{\zeta_h}_{k+1}$, the argmax of the PPO-Clip objective \cite{schulman2017proximal}:
                \begin{equation} 
                \begin{aligned}
                \frac1{BT}\sum_{b=1}^B\sum_{t=0}^T\min\left(\frac{\pi_{\theta^{\zeta_h}}^{\zeta_h}\left(a_t^{\zeta_h}|o_t^{\zeta_h}\right)}{\pi_{\theta_k^{\zeta_h}}^{\zeta_h}\left(a_t^{\zeta_h}|o_t^{\zeta_h}\right)}M^{\zeta_{1:h}}(s_t,\boldsymbol{a}_t), \right. \\
                    \left. \mathrm{clip}\left(\frac{\pi_{\theta^{\zeta_h}}^{\zeta_h}\left(a_t^{\zeta_h}|o_t^{\zeta_h}\right)}{\pi_{\theta_k^{\zeta_h}}^{\zeta_h}\left(a_t^{\zeta_h}|o_t^{\zeta_h}\right)},1\pm\epsilon\right)M^{\zeta_{1:h}}(s_t,\boldsymbol{a}_t)\right);
                \end{aligned}
                \end{equation}
                \STATE Compute unless $\zeta_h$ = $\zeta_V$:
                \begin{equation}
                    M^{\zeta_{1:h+1}}(s,\boldsymbol{a}) = \frac{\pi^{\zeta_h}_{\theta^{\zeta_h}}(a^{\zeta_h}_t \mid o^{\zeta_h}_t)}{\pi^{\zeta_h}_{\theta^{\zeta_h}_k}(a^{\zeta_h}_t \mid o^{\zeta_h}_t)}M^{\zeta_{1:h}}(s, \boldsymbol{a});
                \end{equation}
            \ENDFOR \label{sequence-update-end}
            \STATE Update the V-value network by the following formula:
            \begin{equation}
                \pi_{k+1} = \arg\min_{\pi} \frac{1}{BT} \sum_{b=1}^B \sum_{t=0}^{T_s} \left( V_{\pi_k}(s_t) - \hat{R}_t \right) ^2;
            \end{equation}
        \ENDFOR
    \end{algorithmic}
\end{algorithm}

\section{Experiments}

\begin{figure}[!htbp]
  \centering
  \includegraphics[width=0.48\textwidth]{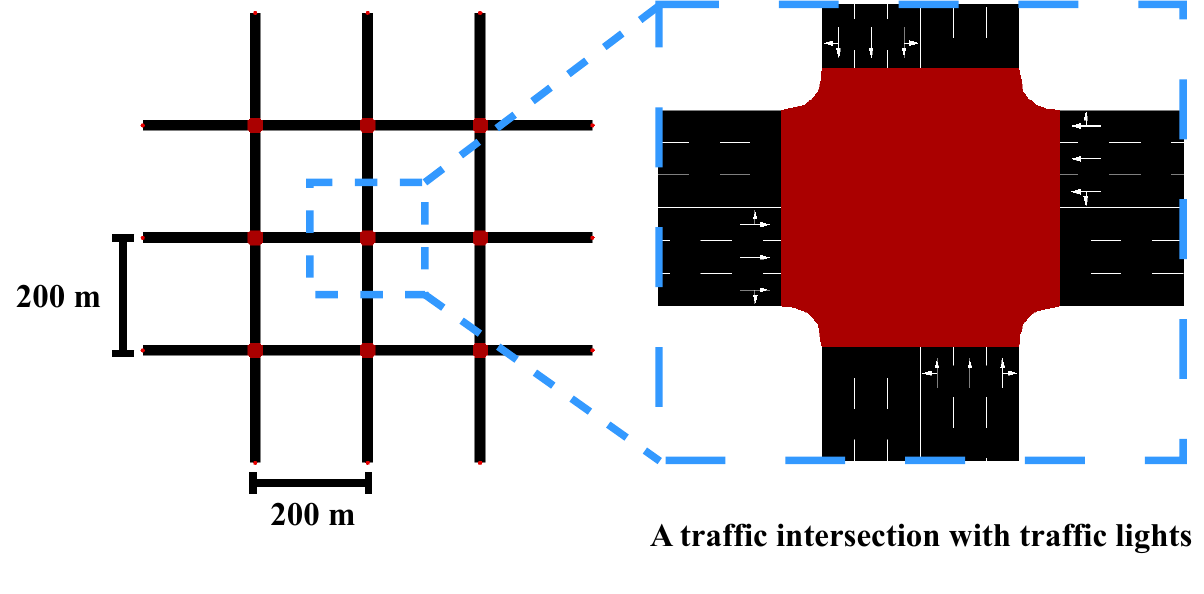}\\
  \caption{The SUMO road network.}
  \label{fig:road}
\end{figure}

Numerical experiments are used to validate the effectiveness of the proposed algorithm. Specifically, we use the SUMO software to generate traffic flow and simulate multi-task FL, as shown in Fig. \ref{fig:road}.
We assume that $H$ vehicles are always driving within the road network and collaboratively complete a total of $M$ tasks.
Each CHV divides the bandwidth into $N$ subcarriers using OFDMA technology.
We consider four classification tasks over datasets MNIST with LeNet \cite{lenet}, Fashion-MNIST \cite{xiao2017fashion} (referred to as FMNIST) with LeNet, SVHN \cite{netzer2011reading} with LeNet, and CIFAR-10 \cite{krizhevsky2009learning} with ResNet-18 \cite{russakovsky2015imagenet}.
Based on the analysis in Section \ref{sec:game-model}, we use the test accuracy to evaluate the model's performance.
Note that under the same experimental setup, different algorithms yield different strategies.
Therefore, the value of $K$ is not fixed across different scenarios or algorithms.
All experiments are conducted using TensorFlow and PyTorch \cite{paszke2019pytorch} with Ubuntu 22.04.
The parameter settings are shown in Tab. \ref{tab:parameter}.

\begin{table}[!htbp]
    \caption{Simulation Parameters \cite{zhang2024joint, yu2024Multi}.}
    \centering
    \begin{tabular}{lc}
        \hline
        Parameters & Values \\ \hline
        Local iterations ($I$) & 5 \\
        Total uplink bandwidth ($W$)  & 20 MHz \\
        Noice power ($\sigma^2$)        & -104 dBm \\
        Channel power gain at 1 m reference distance ($h$) & -34 dBm \\
        Uplink power ($p$)  & 30 dBm \\
        Path loss exponent ($\nu$) & 2 \\
        CPU-cycle frequency ($f$) & 6 GHz \\
        CPU frequency for processing 1 bit of data ($q$) & $10^{3}$ cycles/bit \\
        The effective switching capacitance ($\lambda$) & $10^{-27}$ \\
        Communication radius ($d^U$) & 100 m \\
        The number of vehicles ($H$) & 30 \\
        Maximum time duration ($t^{round}$) & 30 s \\
        The number of subcarriers ($N$) & 60 \\
        Initial vehicle energy & 3,000 J \\
        Distance scaling factor ($\xi$) & 1 \\
        Weighting coefficient ($\varepsilon$) & 1 \\
        Episodes ($K_s$) & 100 \\
        Steps per episode ($T_s$) & 4,000 \\
        Clip range ($\epsilon$) & 0.2 \\
        Discount factor ($\gamma$) & 0.99 \\ \hline
    \end{tabular}
    \label{tab:parameter}
\end{table}
We compare the proposed algorithm with the following three algorithms.
\begin{itemize}
    \item Equal resource allocation (ERA): ERA distributes vehicles evenly across the tasks, randomly designates a leader within the vehicle group of each task, and then evenly allocates subcarriers to the vehicles within the group.
    \item Bayesian optimization-based device scheduling (BODS) \cite{zhou2022efficient}: BODS applies Bayesian optimization for assigning vehicles to the tasks, while the leader selection and bandwidth allocation strategies are the same as in ERA.
    \item Distributed proximal policy optimization (DPPO): DPPO uses deep reinforcement learning to jointly optimize task scheduling, leader selection, and bandwidth allocation; this is a modified version of \cite{zhang2024leader}.
\end{itemize}

\begin{figure}[!htbp]
    \centering
    \includegraphics[width=0.48\textwidth]{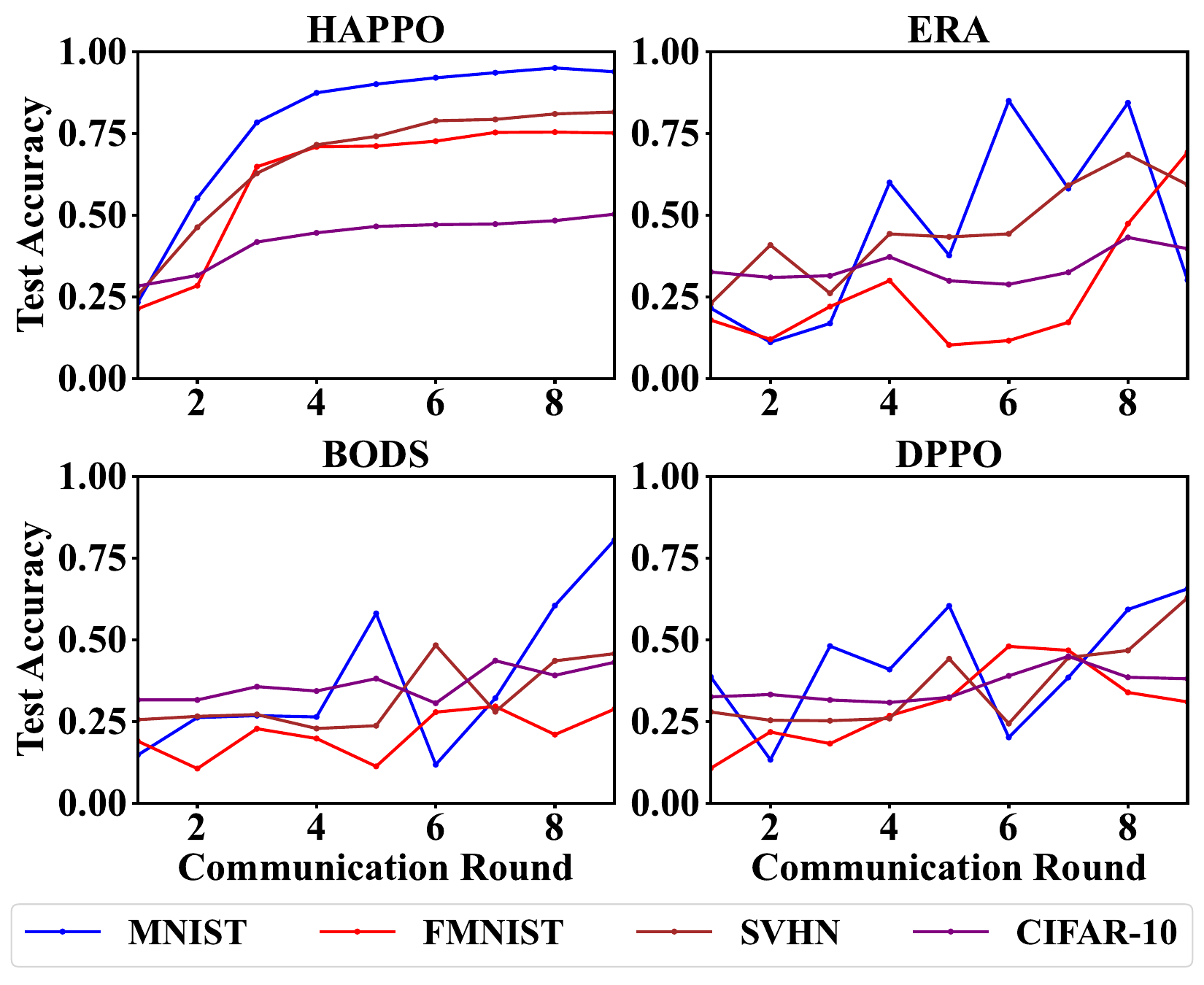}\\
    \caption{Accuracy curves for HAPPO, ERA, BODS, and DPPO.
    }
    \label{fig:acc-curve}
\end{figure}

Fig. \ref{fig:acc-curve} shows the accuracy curves for each task when the algorithms are training on all four tasks simultaneously.
We gain the following insights.
First, the accuracy curve of HAPPO is relatively smooth, indicating that HAPPO can effectively alleviate the issue of model updates being overwritten.
Secondly, the final test accuracy of HAPPO is higher, surpassing the other algorithms by at least 51\%.
This suggests that overwriting model updates indeed can lead to inefficient resource utilization, resulting in a decrease in the final accuracy.
Lastly, the final test accuracy of the four tasks in HAPPO is the highest.
This indicates that HAPPO can utilize resources more efficiently for training.

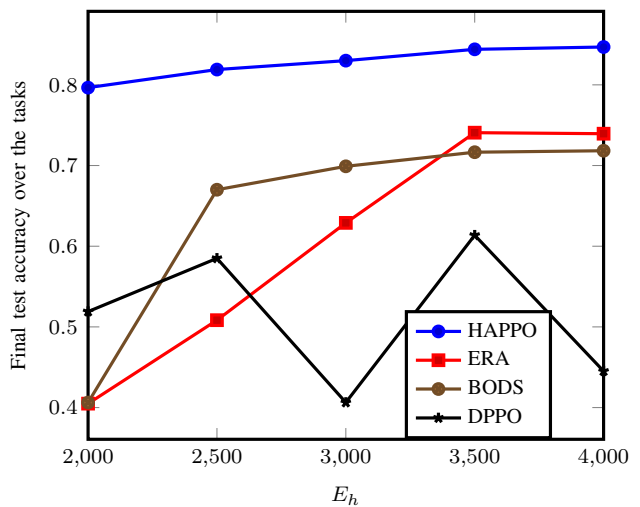
\begin{figure}[!htbp]
    \centering
    \begin{tikzpicture}
    \begin{axis}[  
        xlabel=$E_h$,  
        ylabel=Final test accuracy over the tasks,  
        enlarge x limits=false, 
        legend style={at={(0.90,0.00)},anchor=south east}, 
        legend cell align={left}, 
        mark options={solid}, 
        line width=1.2pt, 
        font=\footnotesize 
    ]  
      
    \addplot table[x=x, y=HAPPO, col sep=comma] {plot/energy_acc.csv};  
    \addlegendentry{HAPPO} %
      
    \addplot table[x=x, y=ERA, col sep=comma] {plot/energy_acc.csv};  
    \addlegendentry{ERA} 
    
    \addplot table[x=x, y=BODS, col sep=comma] {plot/energy_acc.csv};  
    \addlegendentry{BODS}

    \addplot table[x=x, y=DPPO, col sep=comma] {plot/energy_acc.csv};  
    \addlegendentry{DPPO}
      
    \end{axis}  
    \end{tikzpicture}
    \caption{Comparison of final test accuracy with respect to initial vehicle energy level.}
    \label{fig:energy-acc}
\end{figure}

\begin{table}[!htbp]
    \caption{Comparison of maximum time duration over all communication rounds with respect to various scenarios.}
    \centering
    \begin{tabular}{ccccc}
        \hline
        \multirow{2}{*}{Parameter} & \multicolumn{4}{c}{$\max_{k \in \mathcal{K}} T_k$ (s)} \\ \cline{2-5}
        & HAPPO & ERA & BODS & DPPO \\ \hline
        $E_h=2,000$ & 23.8807 & 24.1694 & \textbf{20.9445} & 69.3173 \\
        $E_h=2,500$ & 23.8807 & 24.1694 & \textbf{20.9445} & 69.3173 \\
        $E_h=3,000$ & \textbf{19.3659} & 24.1694 & 20.9445 & 35.4563 \\
        $E_h=3,500$ & 23.3918 & 24.1694 & \textbf{20.9445} & 28.6842 \\
        $E_h=4,000$ & 23.3918 & 24.1694 & \textbf{20.9445} & 31.4053 \\ \hline
        $H=20$ & 15.2720 & 13.6801 & \textbf{11.4227} & 24.9670 \\
        $H=25$ & 21.0747 & 18.8449 & \textbf{15.4588} & 29.0032 \\
        $H=30$ & \textbf{19.3659} & 24.1694 & 20.9445 & 69.3173 \\
        $H=35$ & \textbf{23.8046} & 23.9414 & 23.9414 & 31.2186 \\
        $H=40$ & \textbf{23.7705} & 28.2853 & 28.2853 & 68.9185 \\ \hline
        Task group 1 & \textbf{27.1197} & 46.4546 & 46.4546 & 91.6026 \\
        Task group 2 & \textbf{19.3659} & 23.8807 & 23.8807 & 46.4546 \\
        Task group 3 & \textbf{19.3659} & 24.1694 & 20.9445 & 46.7433 \\
        Task group 4 & 19.3659 & 15.1398 & \textbf{12.8824} & 28.5110 \\
        Task group 5 & 20.7714 & 12.8824 & \textbf{10.6250} & 69.1442 \\ \hline
        $\xi=1.0$ & \textbf{19.3659} & 24.1694 & 20.9445 & 31.4053 \\
        $\xi=1.5$ & 23.9227 & 27.1662 & \textbf{23.5132} & 27.1662 \\
        $\xi=2.0$ & \textbf{25.7886} & 29.8208 & \textbf{25.7886} & 62.9792 \\
        $\xi=2.5$ & \textbf{25.8602} & 32.2873 & 27.9027 & 42.2292 \\
        $\xi=3.0$ & \textbf{27.7431} & 34.6409 & 29.9201 & 100.7320 \\ \hline
    \end{tabular}
    \label{tab:experiment-time}
\end{table}

Fig. \ref{fig:energy-acc} and the first section of Tab. \ref{tab:experiment-time} show the impact of the initial energy level $E_h$ on the algorithms.
We set up three tasks: MNIST with LeNet, FMNIST with LeNet, and SVHN with LeNet.
We gain the following insights.
First, the final test accuracy of HAPPO is higher for all vehicle initial energy levels, surpassing the other algorithms by at least 13\%.
This is because the issue of overwriting continues to affect the entire training process, while HAPPO effectively alleviates this problem, resulting in better performance.
Secondly, as the initial energy of the vehicles increases, the final test accuracy of ERA and BODS also improves.
This is because, as the training process progresses, more vehicles in the framework receive the updated models, which helps to alleviate the issue of overwriting to some extent.
Lastly, DPPO encounters timeout issues frequently.
This is because the large solution space of DPPO leads to incorrect resource allocation strategies, resulting in excessively long transmission times.

\begin{figure}[!htbp]
    \centering
    \begin{tikzpicture}
    \begin{axis}[  
        xlabel=$H$,  
        ylabel=Final test accuracy over the tasks,  
        enlarge x limits=false, 
        legend style={at={(0.50,0.00)},anchor=south east}, 
        legend cell align={left}, 
        mark options={solid}, 
        line width=1.2pt, 
        font=\footnotesize 
    ]  
      
    \addplot table[x=x, y=HAPPO, col sep=comma] {plot/vehicle_acc.csv};  
    \addlegendentry{HAPPO} %
      
    \addplot table[x=x, y=ERA, col sep=comma] {plot/vehicle_acc.csv};  
    \addlegendentry{ERA} 
    
    \addplot table[x=x, y=BODS, col sep=comma] {plot/vehicle_acc.csv};  
    \addlegendentry{BODS}

    \addplot table[x=x, y=DPPO, col sep=comma] {plot/vehicle_acc.csv};  
    \addlegendentry{DPPO}
      
    \end{axis}  
    \end{tikzpicture}
    \caption{Comparison of final test accuracy with respect to the number of vehicles.}
    \label{fig:vehicle-acc}
\end{figure}
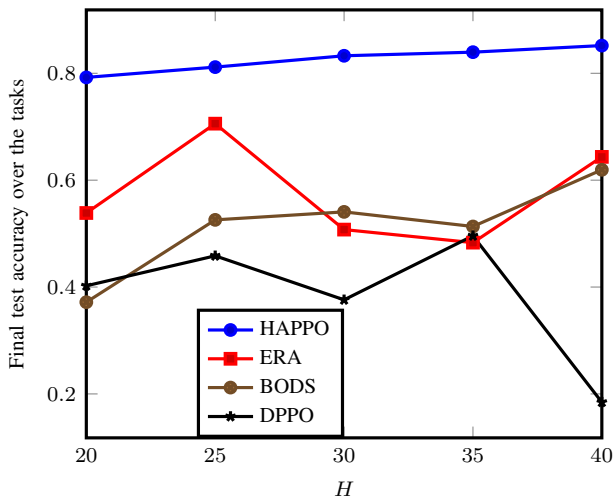

Fig. \ref{fig:vehicle-acc} and the second section of Tab. \ref{tab:experiment-time} show the impact of the number of vehicles $H$ on the algorithms.
We set up three tasks: MNIST with LeNet, FMNIST with LeNet, and SVHN with LeNet.
We gain the following insights.
First, the final test accuracy of HAPPO is higher for all numbers of vehicles, surpassing the other algorithms by at least 14\%.
The increase in the number of vehicles implies an expansion of the solution space, and HAPPO's multi-agent sequential updates can effectively utilize this situation.
Second, the final test accuracy of ERA, BODS, and DPPO exhibits significant fluctuations when varying the number of vehicles.
This is due to the expansion of the solution space, which exacerbates the interference of model updates. Notably, when $H \geq 30$, DPPO experiences timeout.
Lastly, HAPPO demonstrates superior performance in managing the time limit, even as the number of vehicles increases, with at least a 16\% reduction compared to ERA and BODS when $H=40$.
This is because, as the number of vehicles grows, the strategy of uniformly allocating subcarriers becomes less effective. 
Efficient subcarrier allocation based on communication status is needed.
The proposed algorithm efficiently allocates subcarriers, enabling the vehicles to participate in training while avoiding timeout.

\begin{table}[!htbp]
    \caption{Comparison of final test accuracy with respect to different task groups.}
    \centering
    \begin{tabular}{cccccc}
        \hline
        \multirow{2}{*}{Task group} & \multirow{2}{*}{Dataset} & \multicolumn{4}{c}{Final test accuracy} \\ \cline{3-6}
        & & HAPPO & ERA & BODS & DPPO \\ \hline
        1 & MNIST & \textbf{0.9548} & 0.3024 & 0.3298 & 0.4773 \\ \hline
        \multirow{2}{*}{2} & MNIST & \textbf{0.9624} & 0.7728 & 0.8705 & 0.6920 \\
         & FMNIST & \textbf{0.7369} & 0.6133 & 0.7209 & 0.1784 \\ \hline
         \multirow{3}{*}{3} & MNIST & \textbf{0.9524} & 0.3176 & 0.8591 & 0.8956 \\
         & FMNIST & \textbf{0.7707} & 0.6100 & 0.1227 & 0.1213 \\
         & SVHN & \textbf{0.7682} & 0.6085 & 0.6176 & 0.6022 \\ \hline
         \multirow{4}{*}{4} & MNIST & \textbf{0.9388} & 0.3015 & 0.8047 & 0.6553 \\
         & FMNIST & \textbf{0.7514} & 0.6925 & 0.2873 & 0.3093 \\
         & SVHN & \textbf{0.8158} & 0.5934 & 0.4569 & 0.6279 \\
         & CIFAR-10 & \textbf{0.5032} & 0.3977 & 0.4307 & 0.3800 \\ \hline
         \multirow{5}{*}{5} & MNIST & \textbf{0.9336} & 0.1333 & 0.8792 & 0.2813 \\
         & FMNIST & \textbf{0.7220} & 0.2293 & 0.3016 & 0.4800 \\
         & SVHN & \textbf{0.7865} & 0.4690 & 0.2323 & 0.7385 \\
         & CIFAR-10 & \textbf{0.5180} & 0.4620 & 0.4040 & 0.3927 \\
         & CIFAR-10 & \textbf{0.5107} & 0.2926 & 0.4168 & 0.3793 \\ \hline
    \end{tabular}
    \label{tab:task-acc}
\end{table}

Tab. \ref{tab:task-acc} and the third section of Tab. \ref{tab:experiment-time} show the impact of the task group on the algorithms.
Please note that in Task group 5, there are two tasks involving CIFAR-10 with ResNet-18, and these two tasks are trained independently.
We gain the following insights.
Firstly, HAPPO outperforms the other algorithms in terms of accuracy across all task combinations. By dynamically adjusting the action space, HAPPO is particularly well-suited for multi-task FL, effectively handling different task combinations.
Secondly, ERA, BODS, and DPPO fail when $M=1$, while HAPPO exhibits strong performance. This is because when $M=1$, the task scheduling problem reduces to a client selection problem. ERA and BODS are unable to effectively determine which clients should participate, resulting in excessive bandwidth pressure.
DPPO continues to experience timeout issues due to the large solution space.
Lastly, when $M \leq 3$, HAPPO requires less time within a round. As the number of tasks increases, this time required by ERA and BODS also decreases, but their final test accuracy does not improve steadily.
When the number of tasks is small, each task experiences higher bandwidth pressure.
The proposed algorithm effectively allocates subcarriers to mitigate this pressure. 
As the number of tasks increases, bandwidth pressure decreases, making leader selection management important for mitigating the issue of overwriting.

\begin{figure}[!htbp]
    \centering
    \begin{tikzpicture}
    \begin{axis}[  
        xlabel=$\xi$,  
        ylabel=Final test accuracy over the tasks,  
        enlarge x limits=false, 
        legend style={at={(0.40,0.00)},anchor=south east}, 
        legend cell align={left}, 
        mark options={solid}, 
        line width=1.2pt, 
        font=\footnotesize 
    ]  
      
    \addplot table[x=x, y=HAPPO, col sep=comma] {plot/xi_acc.csv};  
    \addlegendentry{HAPPO} %
      
    \addplot table[x=x, y=ERA, col sep=comma] {plot/xi_acc.csv};  
    \addlegendentry{ERA} 
    
    \addplot table[x=x, y=BODS, col sep=comma] {plot/xi_acc.csv};  
    \addlegendentry{BODS}

    \addplot table[x=x, y=DPPO, col sep=comma] {plot/xi_acc.csv};  
    \addlegendentry{DPPO}
      
    \end{axis}  
    \end{tikzpicture}
    \caption{Comparison of final test accuracy with respect to the distance scaling factor.}
    \label{fig:xi-acc}
\end{figure}
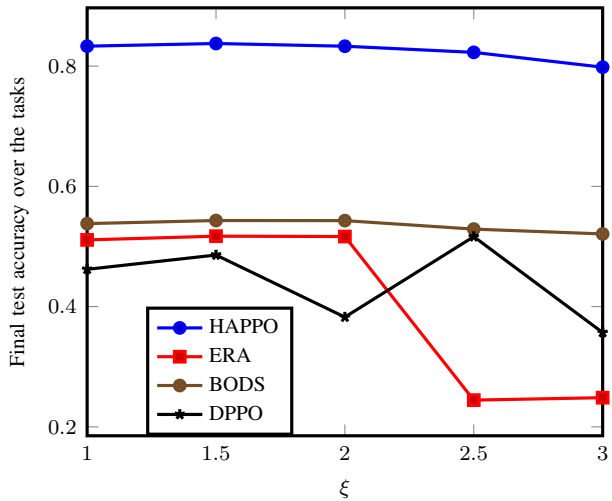

Fig. \ref{fig:xi-acc} and the fourth section of Tab. \ref{tab:experiment-time} show the impact of 
the distance scaling factor $\xi$.
We set up three tasks: MNIST with LeNet, FMNIST with LeNet, and SVHN with LeNet.
We gain the following insights.
The final test accuracy of HAPPO is highest among all four algorithms, with a margin of at least 53\%, though a downward trend is observed.
This is because the indirect communication cost increases, leading to a reduction in the number of vehicles participating in the training.
In addition, the increase in communication cost is also reflected in $T_k$. The $T_k$ of HAPPO, ERA, and BODS increases as the communication cost rises. ERA experiences a timeout when $\xi \geq 2.5$.



\section{Conclusion}

In this paper, we have proposed a mobility-aware multi-task decentralized federated learning framework for vehicular networks.
Considering the impact of mobility and resource constraints, we have presented a joint optimization problem for task scheduling, subcarrier allocation, and leader selection.
For problem solving, we analyze the convergence bound of a single FL task, and then model multiple FL tasks as a resource allocation game.
The game is further reformulated as a DEC-POMDP, and we propose an HAPPO-based algorithm to solve it.
We compare the proposed algorithm with baselines under varying initial vehicle energy, number of vehicles, number of tasks, and indirect communication cost.
We obtain the following three main insights.
First, the experiment results show that the proposed algorithm improves the final average accuracy by at least 13\%.
Second, for all the experiments, the proposed algorithm efficiently completes the training, while ERA and BODS both encounter time-out in some cases.
These experimental results demonstrate that the proposed algorithm can stably solve the TSLP problem.
Finally, the proposed algorithm exhibits more stable convergence compared to DPPO.
These experimental results indicate that the proposed algorithm can search for the NE solution more efficiently.

\appendices

\section{Proof of Lemma 1}
\label{appendix-a}
According to Assumption 1 and the definition of the global loss in (\ref{eq:global-loss}), The global loss function also has the $L$-smooth property:
\begin{equation}
    \begin{aligned}
        & F^m(\boldsymbol{w}^m_k)-F^m(\boldsymbol{w}^m_{k-1}) \\
        & \leq
        \langle \nabla F^m(\boldsymbol{w}^m_{k-1}),\boldsymbol{w}^m_k - \boldsymbol{w}^m_{k-1} \rangle
        + \frac{L}{2} \left\|\boldsymbol{w}^m_k-\boldsymbol{w}^m_{k-1}\right\|^2.
    \end{aligned}
    \label{eq:L-smooth-global-loss}
\end{equation}

First, we introduce Lemma 2.

\textbf{Lemma 2.} \textit{The magnitude of the global gradient descent can be defined by the local updates of the clients.}

\begin{proof}
For the term $\mathbb{E}\left[
    \frac
    {\sum_{h\in\mathcal{H}}\alpha^m_{kh}|\mathcal{D}^m_h|(\boldsymbol{w}^m_{kh}-\boldsymbol{w}^m_{k-1})}
    {\sum_{h\in\mathcal{H}}\alpha^m_{kh}|\mathcal{D}^m_h|}
    \right]$, we have
\begin{equation}
    \begin{aligned}
    & \mathbb{E}\left[
    \frac
    {\sum_{h\in\mathcal{H}}\alpha^m_{kh}|\mathcal{D}^m_h|(\boldsymbol{w}^m_{kh}-\boldsymbol{w}^m_{k-1})}
    {\sum_{h\in\mathcal{H}}\alpha^m_{kh}|\mathcal{D}^m_h|}
    \right] \\
    &= 
    \mathbb{E}\left[ -
    \frac
    {\sum_{h\in\mathcal{H}}\sum_{x\in\mathcal{B}_{kh}}\alpha^m_{kh}|\mathcal{D}^m_h| \frac{\eta_k}{B_{k}} \nabla f^m(\boldsymbol{w}^m_{k-1};x)}
    {\sum_{h\in\mathcal{H}}\alpha^m_{kh}|\mathcal{D}^m_h|}
    \right] \\
    &= 
    \mathbb{E}\left[ - \eta_k
    \frac
    {\sum_{h\in\mathcal{H}}\alpha^m_{kh}|\mathcal{D}^m_h| \nabla f^m_h(\boldsymbol{w}^m_{k-1})}
    {\sum_{h\in\mathcal{H}}\alpha^m_{kh}|\mathcal{D}^m_h|}
    \right] \\
    &= - \eta_k \nabla F^m ( \boldsymbol{w}^m_{k-1} ),
    \end{aligned}
    \label{eq:lemma-2}
\end{equation}
where the first equality holds according to the local model gradient descent formula in (\ref{eq:loss-define}); the second and third equalities hold according to Assumption 3.
\end{proof}

Next, we estimate the expectation of Equation (\ref{eq:L-smooth-global-loss}). For the first term of (\ref{eq:L-smooth-global-loss}), we have
\begin{equation}
    \begin{aligned}
    & \mathbb{E} \left[ \langle \nabla F^m(\boldsymbol{w}^m_{k-1}),\boldsymbol{w}^m_k - \boldsymbol{w}^m_{k-1} \rangle \right] \\
    &= \langle \nabla F^m(\boldsymbol{w}^m_{k-1}), \mathbb{E} \left[ \boldsymbol{w}^m_k - \boldsymbol{w}^m_{k-1} \right] \rangle \\
    &= \langle \nabla F^m(\boldsymbol{w}^m_{k-1}), \mathbb{E} \left[  \frac{\sum_{h\in\mathcal{H}}\alpha^m_{kh}|\mathcal{D}^m_h|(\boldsymbol{w}^m_{kh}-\boldsymbol{w}^m_{k-1})}{\sum_{h\in\mathcal{H}}\alpha^m_{kh}|\mathcal{D}^m_h|} \right] \rangle \\
    &= \langle \nabla F^m(\boldsymbol{w}^m_{k-1}), - \eta_k \nabla F^m(\boldsymbol{w}^m_{k-1}) \rangle  \\
    &= - \eta_k \left\|{\nabla}F^m(\boldsymbol{w}^m_{k-1})\right\|^2 ,
    \end{aligned}
    \label{eq:first-term-lemma-1}
\end{equation}
where the third equality holds according to Lemma 2.

For the second term of (\ref{eq:L-smooth-global-loss}), we have
\begin{equation}
    \begin{aligned}
    & \mathbb{E} \left[ \left\|\boldsymbol{w}^m_k-\boldsymbol{w}^m_{k-1}\right\|^2 \right] \\
    &= \mathbb{E} \left[ \left\| \frac
    {\sum_{h\in\mathcal{H}}\alpha^m_{kh}|\mathcal{D}^m_h|(\boldsymbol{w}^m_{kh}-\boldsymbol{w}^m_{k-1})}
    {\sum_{h\in\mathcal{H}}\alpha^m_{kh}|\mathcal{D}^m_h|} \right\|^2 \right] \\
    &= \mathbb{E} \left[ \left\| - \frac{\eta_k}{B_{k}}
    \frac
    {\sum_{h\in\mathcal{H}}\sum_{x\in\mathcal{B}_{kh}}\alpha^m_{kh}|\mathcal{D}^m_h| \nabla f^m(\boldsymbol{w}^m_{k-1};x)}
    {\sum_{h\in\mathcal{H}}\alpha^m_{kh}|\mathcal{D}^m_h|} \right\|^2 \right] \\
    &\leq \eta_k^2 \left( \left\|
    \nabla F^m(\boldsymbol{w}^m_{k-1})
    \right\|^2 + \frac{G^2}{B_{k}\sum_{h\in\mathcal{H}}\alpha^m_{kh}} \right),
    \end{aligned}
    \label{eq:second-term-lemma-1}
\end{equation}
where the first and second equalities hold according to the local model gradient descent formula; the third inequality holds according to Assumption 4.

Taking the expectation of both sides of (\ref{eq:L-smooth-global-loss}) and substituting (\ref{eq:first-term-lemma-1}) and (\ref{eq:second-term-lemma-1}), we have
\begin{equation}
    \begin{aligned}
    & \mathbb{E} \left[ F^m(\boldsymbol{w}^m_k) \right] - \mathbb{E} \left[ F^m(\boldsymbol{w}^m_{k-1}) \right] \\
    & \leq - \eta_k\left\|{\nabla}F^m(\boldsymbol{w}^m_{k-1})\right\|^2 \\
    & + \frac{L\eta_k^2}{2} \left( \left\|
    \nabla F^m(\boldsymbol{w}^m_{k-1})
    \right\|^2 + \frac{G^2}{B_{k}\sum_{h\in\mathcal{H}}\alpha^m_{kh}} \right) \\
    & = \eta_k \left( \frac{L\eta_k}{2} - 1 \right)\left\|
    \nabla F^m(\boldsymbol{w}^m_{k-1})
    \right\|^2 + \frac{L\eta_k^2}{2} \frac{G^2}{B_{k}\sum_{h\in\mathcal{H}}\alpha^m_{kh}}.
    \end{aligned}
\end{equation}
Lemma 1 is proved.
 $\hfill\square$

\section{Proof of Theorem 1}
\label{appendix-b}
According to Assumption 2, the Polyak-Lojasiewicz inequality holds
\begin{equation}
    \| \nabla F^m(\boldsymbol{w}^m_{k-1})^2 \| \geq 2 \mu \left( F^m(\boldsymbol{w}^m_{k-1})-F^m(\boldsymbol{w}^{m*}) \right).
    \label{PL}
\end{equation}

Assuming $\eta_k \leq \frac{1}{L}$, we have
\begin{equation}
    \begin{aligned}
    &\mathbb{E} \left[ F^m(\boldsymbol{w}^m_k) \right] - \mathbb{E} \left[ F^m(\boldsymbol{w}^m_{k-1}) \right] \\
    & \leq \eta_k \left( \frac{L\eta_k}{2} - 1 \right)\left\|
    \nabla F^m(\boldsymbol{w}^m_{k-1})
    \right\|^2 + \frac{L\eta_k^2}{2} \frac{G^2}{B_{k}\sum_{h\in\mathcal{H}}\alpha^m_{kh}} \\
    & \leq -\frac{\eta_k}{2} \left\|
    \nabla F^m(\boldsymbol{w}^m_{k-1})
    \right\|^2 + \frac{\eta_k}{2} \frac{G^2}{B_{k}\sum_{h\in\mathcal{H}}\alpha^m_{kh}} \\
    & \leq -\eta_k\mu \left( \mathbb{E} [ F^m(\boldsymbol{w}^m_{k-1})-F^m(\boldsymbol{w}^{m*}) ] \right) + \frac{\eta_k}{2} \frac{G^2}{B_{k}\sum_{h\in\mathcal{H}}\alpha^m_{kh}},
    \end{aligned}
    \label{eq:theorem-1-base}
\end{equation}
where the first inequality holds according to Lemma 1, the second inequality holds according to the specified learning rate condition, and the third inequality holds according to (\ref{PL}).

Rearranging the left-hand side of (\ref{eq:theorem-1-base}) and applying recursion to the equation, we have
\begin{equation}
    \begin{aligned}
    & \mathbb{E} \left[ F^m(\boldsymbol{w}^m_K) \right] - F^{m}(\boldsymbol{w}^{m*}) \\
    & \leq (1 - \mu\eta_K) \left( \mathbb{E} [ F^m(\boldsymbol{w}^m_{K-1}) ] -F^m(\boldsymbol{w}^{m*}) \right) \\
    & + \frac{\eta_K}{2} \frac{G^2}{B_{K}\sum_{h\in\mathcal{H}}\alpha^m_{Kh}} \\
    & \leq \left( \mathbb{E} [ F^m(\boldsymbol{w}^m_{0}) ] -F^m(\boldsymbol{w}^{m*}) \right) \prod_{k=1}^{K}(1-\mu\eta_{k}) \\
    & + \sum_{k=1}^{K-1}\frac{\eta_k}{2}\frac{G^2}{B_{k}\sum_{h\in\mathcal{H}}\alpha^m_{kh}} \prod_{j=k+1}^{K}(1-\mu\eta_{j}) \\
    & + \frac{\eta_K}{2}\frac{G^2}{B_{K}\sum_{h\in\mathcal{H}}\alpha^m_{Kh}}.
    \end{aligned}
    \nonumber
\end{equation}
Theorem 1 is proved.
 $\hfill\square$

\section{Proof of Theorem 2}
\label{appendix-c}
According to (\ref{eq:def-2}) and (\ref{eq:theorem-2}), we have
\begin{equation}
    \begin{aligned}
        & \Omega^m_k(\mathscr{S}^{m{\prime}}_k,\mathscr{S}^{-m}_k)-\Omega^m_k(\mathscr{S}^m_k,\mathscr{S}^{-m}_k) \\
        &= \sum_{i \in \mathcal{M}} \left[ U^i_k(\mathscr{S}^{m{\prime}}_k,\mathscr{S}^{-m}_k)-U^i_k(-\mathscr{S}_k^{m\prime},\mathscr{S}^{-m}_k) \right] \\
        &\quad - \sum_{i \in \mathcal{M}} \left[ U^i_k(\mathscr{S}^m_k,\mathscr{S}^{-m}_k)-U^i_k(-\mathscr{S}^m_k,\mathscr{S}^{-m}_k) \right] \\
        &= \sum_{i \in \mathcal{M}} \left[ U^i_k(\mathscr{S}^{m{\prime}}_k,\mathscr{S}^{-m}_k) - U^i_k(\mathscr{S}^m_k,\mathscr{S}^{-m}_k) \right. \\
        &\quad + \left. U^i_k(-\mathscr{S}^m_k,\mathscr{S}^{-m}_k) - U^i_k(-\mathscr{S}_k^{m\prime},\mathscr{S}^{-m}_k) \right] \\
        &= \sum_{i \in \mathcal{M}} \left[ U^i_k(\mathscr{S}^{m{\prime}}_k,\mathscr{S}^{-m}_k) - U^i_k(\mathscr{S}^m_k,\mathscr{S}^{-m}_k) \right] \\
        &= \beta^m \log(\frac{\sum_{h \in \mathcal{H}}\alpha^{m\prime}_{kh}}{\sum_{h \in \mathcal{H}}\alpha^m_{kh}}).
    \end{aligned}
    \nonumber
\end{equation}
According to (\ref{eq:def-3}) in Definition 3, assuming $U^m_k(\mathscr{S}^{m{\prime}}_k,\mathscr{S}^{-m}_k) > U^m_k(\mathscr{S}^m_k,\mathscr{S}^{-m}_k)$, we have
\begin{equation}
    \psi^{m\prime}_k > \psi^m_k.
    \nonumber
\end{equation}
By monotonicity of (\ref{eq:def-1}), we have
\begin{equation}
    \begin{aligned}
        & \sum_{h \in \mathcal{H}}\alpha^{m\prime}_{kh} > \sum_{h \in \mathcal{H}}\alpha^{m}_{kh}, \\
        & \log(\frac{\sum_{h \in \mathcal{H}}\alpha^{m\prime}_{kh}}{\sum_{h \in \mathcal{H}}\alpha^m_{kh}}) > 0.
    \end{aligned}
    \nonumber
\end{equation}
Together with $\beta^m > 0$, we have
\begin{equation}
    \Omega^m_k(\mathscr{S}^{m{\prime}}_k,\mathscr{S}^{-m}_k)-\Omega^m_k(\mathscr{S}^m_k,\mathscr{S}^{-m}_k) > 0.
    \nonumber
\end{equation}
According to Definition 3, the game $\mathcal{G}_k$ is a potential game, the proof of the theorem is complete.
$\hfill\square$

\makeatletter
\patchcmd{\@IEEEeqnarraynumpages}%
  {\def\@IEEEeqnarraydigits{\,}}
  {\def\@IEEEeqnarraydigits{}}
  {}{}
\makeatother

\bibliographystyle{IEEEtran}
\bibliography{ref}


 
\vspace{11pt}

\vspace{11pt}


\vfill

\end{document}